\documentclass[10pt, conference, compsocconf]{IEEEtran}
\usepackage{cite}
\usepackage{amsmath,amssymb,amsfonts,amsthm}
\usepackage{algorithmic}
\usepackage{graphicx}
\usepackage{textcomp}
\usepackage{xcolor}

\usepackage{footnote}
\usepackage{caption}
\usepackage{listings}
\usepackage{framed,multicol}
\usepackage{tabularx}
\usepackage{booktabs}
\usepackage{multirow}
\usepackage{threeparttable}
\usepackage{siunitx}
\usepackage{ragged2e}
\usepackage{subcaption}
\newcommand{\red}{\textcolor{black}}

\usepackage{hyperref}

\usepackage{nomencl}
\usepackage{tikz}
\newcommand*\emptycirc[1][1ex]{\tikz\draw (0,0) circle (#1);} 

\newcommand*\fullcirc[1][1ex]{\tikz\fill (0,0) circle (#1);} 

\usepackage[
	n,
	operators,
	advantage,
	sets,
	adversary,
	landau,
	probability,
	notions,	
	logic,
	ff,
	mm,
	primitives,
	events,
	complexity,
	asymptotics,
	keys]{cryptocode}

\newcommand{\pchline}{\\[-0.5\baselineskip][\hline] \\[-0.5 \baselineskip]}

\newcommand{\pccolone}[1]{#1 \< \<}
\newcommand{\pccoltwo}[1]{\< #1 \<}
\newcommand{\pccolthr}[1]{\< \< #1}

\newcommand{\sffont}[1]{\ensuremath{\mathsf{#1}}}
\newcommand{\func}[1]{\sffont{#1}}
\newcommand{\prp}{\func{PRP}}
\newcommand{\pr}{\func{PRF}}

\newcommand{\brown}[1]{\textcolor{black}{#1}}
\newcommand{\blue}[1]{\textcolor{black}{#1}}
\newcommand{\green}[1]{\textcolor{black}{#1}}

\newcommand{\violet}[1]{\textcolor{black}{#1}}	

\newenvironment{proto-func}[3]{
\def \protocap {#2} 
\def \protolab {#3}
\begin{figure*}
\begin{framed}
\begin{multicols}{2}
\footnotesize
\begin{center}
\textbf{#1}
\end{center}}
{\end{multicols}
\vspace{-10pt}
\end{framed}
\caption{\protocap}
\label{\protolab}
\vspace{-10pt}
\end{figure*}
}

\newenvironment{pevent}[2]{
\begin{flushleft}
\textbf{#1}
{#2}
\end{flushleft}
\begin{flushleft}
}
{
\end{flushleft}	
}

\newcommand{\psmallindent}[1][8]{\par}
\newcommand{\pindent}[1][8]{\par\hspace{#1pt}}
\newcommand{\pif}{\textbf{if }}
\newcommand{\pelse}{\textbf{else }}

\newcommand{\pcomment}[1]{\hfill // #1}

\newcommand{\var}[1]{\ensuremath{\mathit{#1}}}

\newcommand{\msg}[1]{\text{\blue{``#1''}}}
\newcommand{\party}[2][]{\brown{\ensuremath{\mathcal{#2}_{#1}}}}
\newcommand{\assign}{\coloneqq}
\newcommand{\money}[1]{\ensuremath{\green{\var{\$#1}}}}

\newcommand{\const}[1]{\text{\uppercase{#1}}}

\theoremstyle{plain}
\newtheorem{thm}{Theorem} 

\theoremstyle{definition}
\newtheorem{defn}{Definition} 

\begin{document}
\title{Towards Privacy-assured and Lightweight On-chain Auditing of Decentralized Storage}

\author{\IEEEauthorblockN{Yuefeng Du\IEEEauthorrefmark{1}\IEEEauthorrefmark{3}, Huayi Duan\IEEEauthorrefmark{1}\IEEEauthorrefmark{3}, Anxin Zhou\IEEEauthorrefmark{1}\IEEEauthorrefmark{3}, Cong Wang\IEEEauthorrefmark{1}\IEEEauthorrefmark{3}, Man Ho Au\IEEEauthorrefmark{2}, and Qian Wang\IEEEauthorrefmark{4}}
\IEEEauthorblockA{\IEEEauthorrefmark{1}\textit{City University of Hong Kong}, Hong Kong; \IEEEauthorrefmark{2}\textit{The University of Hong Kong}, Hong Kong}\IEEEauthorblockA{\IEEEauthorrefmark{3}\textit{CityU University of Hong Kong Shenzhen Research Institute}, China; \IEEEauthorrefmark{4}\textit{Wuhan University, China}}
}

\maketitle

\noindent\textbf{This paper will be presented at 40th IEEE International Conference on Distributed Computing Systems (ICDCS 2020). Please cite this work as:}

\noindent\textit{Yuefeng Du, Huayi Duan, Anxin Zhou, Cong Wang, Man Ho Au, Qian Wang, ``Towards Privacy-assured and Lightweight On-chain Auditing of Decentralized Storage'' in Proceedings of the 40th IEEE International Conference on Distributed Computing Systems, Singapore, November 2020.}
\newline
\\

\begin{abstract}
How to audit outsourced data in centralized storage like cloud is well-studied, but it is largely under-explored for the rising decentralized storage network (DSN) that bodes well for a billion-dollar market. To realize DSN as a usable service in a fully decentralized manner, the blockchain comes in handy --- to record and verify audit trails in forms of proof of storage, and based on that, to enforce fair payments with necessary dispute resolution.

Leaving the audit trails on the blockchain offers transparency and fairness, yet it 1) sacrifices privacy, as they may leak information about the data under audit, and 2) overwhelms on- chain resources, as they may be practically large in size and expensive to verify. Prior auditing designs in centralized settings are not directly applicable here. A handful of proposals targeting DSN cannot satisfactorily address these issues either.

We present an auditing solution that addresses on-chain privacy and efficiency, from a synergy of homomorphic linear authenticators with polynomial commitments for succinct proofs, and the sigma protocol for provable privacy. The solution results in, per audit, 288-byte proof written to the blockchain, and constant verification cost. It can sustain long-term operation and easily scale to thousands of users on Ethereum.

\end{abstract}

\IEEEpeerreviewmaketitle

\section{Introduction}
\label{sec-introduction}

Within the past few decades, the cloud storage providers have reshaped the data storage marketplace remarkably.
Though the business model of centralized data storage has boomed related industries, the downside of centralism is that this mode poses new security and privacy threats upon user data.
The cloud storage providers could peek at or even abuse user data, as users essentially lose data custodian when they outsource sensitive personal data.
Moreover, centralism of data storage profits merely a few storage providers in the cloud era, leading to the phenomenon of market monopoly.

Alternatively, volunteer-based P2P storage systems, such as Bittorrent~\cite{cohen2003incentives} and IPFS~\cite{benet2014ipfs}, has pointed out a new direction for data storage since the early 2000s.
However, the robustness and reliability of P2P storage systems could be a problem in practice, as they are often abused for piracy and illegal content sharing.
The profound reason is that in such a system built on volunteers, peer nodes may join and leave freely, without ever worrying about the influence of either good or bad behaviours.

Fortunately, the take-off of blockchain technology~\cite{nakamoto2008bitcoin,wood2014ethereum} has risen conclusively and has been still rapidly evolving.
By harnessing the power of the blockchain, it demonstrates a new approach to create an ecosystem of decentralized storage with an incentive mechanism. 
%
%
Imagine we could have our P2P storage system incentively enabled by the blockchain.
On that account, it would be possible to store personal data on such an incentive-driven decentralized cloud storage system.

With a blockchain-enabled decentralized storage system, the users would obtain stronger security guarantees in the sense that the storage providers cannot learn data contents encrypted at the user end.
In addition, this system naturally provides an opportunity to exploit tremendous amounts of underutilized storage space and further benefit the public at large instead of merely profiting a few cloud storage corporations.
%

It is worth mentioning that such a blockchain-enabled decentralized storage system should not be regarded as a replacement, but as a complement of the current cloud storage services.
Notably, it is even likely for the storage \red{providers} to utilize the existing cloud-based CDN infrastructures, e.g., outsource to the cloud (as long as data is encrypted), to provide high-quality service of caching and data retrieval.
Under this decentralized storage system, where the blockchain serves as the backbone incentivizing system, we vision to offer a data storage option benefiting all participants, especially for users with sensitive data that requires stronger security guarantees and storage providers with unused storage space to rent out.

\subsection{Target problem}
Consider a scenario where a user intends to back up his or her photo collections off-site.
Though the user does not fully trust the cloud storage providers, the entrust to unknown storage providers in the decentralized context seems even more unreliable.
Even with a robust incentive system that can punish the wrongdoers, the user may never find out whether partial data is lost until the time of data retrieval, which is utterly inconvenient and costly.
On the other hand, even if we assume the user successfully captures the wrongdoers, how could the blockchain assist in resolving this dispute, and what evidence should serve as the sole and final judging criterion?

To summarize the core problem, we lack a effective yet low-cost storage auditing system and also a fair dispute resolution mechanism for the blockchain-enabled decentralized storage system.
Straightforwardly, in the pursuit of lightweight auditing, the participants can depend on off-chain auditing.
Nonetheless, it introduces additional trust assumptions and eventually requires the blockchain to resolve disputes.
Therefore, as a starting effort, we assume that all of auditing work is carried out with on-chain functionalities.
%

\noindent\textbf{Challenges.} Designing and deploying a workable prototype could be challenging due to the following factors.
Firstly, we have to deal with the tension between the privacy concerns and the transparency property of the blockchain.
According to the recent interpretation of the GDPR~\cite{finck2018blockchains,GDPRpositionpaper}, it is recommended to limit the amount of personal data (even encrypted) on the blockchain in case of future breaches.
In the case of blockchain-enabled auditing, putting insecure proofs onto the blockchain may give the adversaries an opportunity to recover the original data contents off-line in a brutal-force manner, especially when standard deduplication techniques (deterministic encryption) are commonly applied to storage services.
Secondly, aside from the privacy concerns, \red{another challenge} is to design the auditing protocol with succinct proofs and quick verification.
This matters significantly to the cost-effectiveness and even the scalability of the entire decentralized storage system, considering the cost on the blockchain amortized to all miners.
In addition, minimizing the amount of work on the side of data owner as well as the overhead brought to the storage provider is also vital to the practical deployment of our system.

\noindent\textbf{Remarks.} As a starting effort, we focus on the specific application case of personal and enterprise archive data storage.
Once data is distributed and archived, there would be no more update of data.
The market of archive storage alone boasts great potential, including some of the most popular usage scenarios such as file collection archiving and image backups.
From another perspective, most blockchain-enabled storage systems, e.g., Storj~\cite{storjstat}, Siacoin~\cite{siacoin2014}, Filecoin~\cite{filecoin2017}, do not support dynamic data update.
Our proposed design would undoubtedly be comparable in the settings of decentralized archive data storage.

\subsection{Our contributions}
\label{subsec-intro-contribution}

To the best of our knowledge, our blockchain-enabled auditing framework is the first to address on-chain privacy concerns and on-chain efficiency considerations. Specifically, we make the following contributions.

\begin{itemize}
\item We analyze the current practices of decentralized archive storage systems and identify the key problems that hinder their further development. 
%
%
\item We propose the first pragmatic data auditing protocol with on-chain privacy and efficiency for the decentralized storage paradigm.
\red{Whilst achieving on-chain privacy, we manage to cut down the on-chain auditing cost to merely $0.1$\$ per audit per storage provider}.
\item We conduct comprehensive performance evaluation to show our design also accomplishes off-chain efficiency and would remain reliable and robust, even when scaling up to thousands of users. 
\end{itemize}
To summarize, our design of auditing makes the decentralized archive storage comparable to cloud storage solutions in terms of both cost and effectiveness. 

\section{Background}
\label{DSN-background}

\begin{table}[!t]
\small
\centering
\begin{threeparttable}
\setlength{\tabcolsep}{0.05 em}
\captionsetup{justification=centering}
\caption{Comparison of auditing-related features for frameworks that can support decentralized storage.}
\label{tab:compare}
\begin{tabular}{@{}c|cccccccc@{}}
\toprule
\multirow{2}{*}{} & \multicolumn{2}{c}{w.o. audit} & \multicolumn{3}{c}{w. Merkle tree} & \multicolumn{3}{c}{w. SNARK-based} \\ \cmidrule(l){2-3} \cmidrule(l){4-6} \cmidrule(l){7-9}
                  & IPFS          & Swarm           & Storj       & MaidSafe    & Sia        & Filecoin   & ZKCSP    & Hawk     \\ \toprule
Class             & P2P           & EC              & ALT         & ALT         & ALT        & ALT        & BC       & EC       \\ 
Incentive         & \emptycirc          & \emptycirc             & \fullcirc         & \fullcirc         & \fullcirc        & \fullcirc        & \fullcirc     & \fullcirc      \\ 
Audit mode        & N/A              & TTP                 & TTP               & TTP           & BC               & PA              & PA            & BC           \\ 
Storage Guar.     & N/A        & Low           & Low           & Low       & Low         & High         & High        & N/P       \\
On-chain Sec.     & \emptycirc          & \emptycirc             & \fullcirc         & \fullcirc         & \emptycirc       & \emptycirc       & \fullcirc     & \fullcirc      \\ 
Prover eff.       & \emptycirc          & \fullcirc           & \fullcirc         & \fullcirc         & \fullcirc        & \emptycirc        & \emptycirc       & \emptycirc        \\ 
Auditor eff.      & \emptycirc          & \fullcirc           & \fullcirc         & \fullcirc         & \fullcirc        & \fullcirc        & \fullcirc     & \fullcirc      \\ \bottomrule
\end{tabular}
\begin{tablenotes}[flushleft]
\small
\item \tnote{a} \emptycirc \, indicates the feature is not considered by design; \fullcirc \, means the feature is fully supported by design; N/A stands for non-applicable feature; N/P means the feature may be supported but not specified.
\item \tnote{b} Class: classified according to different categories, P2P or Ethereum-compatible (EC) or Bitcoin-compatible (BC) or Altercoin (ALT). We consider Ethereum-based auditing solution is more universal for its concept of generic smart contracts; Bitcoin-based auditing solution is compatible with Ethereum platform, though a very limited amount of computation can be done on the blockchain; while Altercoin design cannot be used in other cases.
\item \tnote{c} Audit mode: TTP stands for trusted third party; BC indicates blockchain-enable auditing; PA stands for private auditing; Filecoin claims the data owner can act as the auditor~\cite{filecoin2017}.
\item \tnote{d} Storage guarantees: Naive Merkle-tree based auditing can only provide limited storage guarantees, as the challenge randomness would eventually run out and the prover may reuse the challenged blocks; Leveraging the recursive composite SNARK~\cite{filecoin2017}, Filecoin circumvents the above issue; ZKCSP that applies homomorphic authenticators are known to provide high storage guarantees.
\end{tablenotes}
\end{threeparttable}
\vspace{-15pt}
\end{table}

\red{In this section, we perform a thorough analysis on existing DSN designs. See Table~\ref{tab:compare} for an overview.}

Cryptocurrencies like Bitcoin~\cite{nakamoto2008bitcoin} and Ethereum~\cite{wood2014ethereum} that are founded on the blockchain technology enable a way to support storage auditing in a decentralized fashion.
In particular, Ethereum is extremely helpful because it empowers the execution of any highly expressive smart contract (quasi Turing completeness) bounded by the gas cost.
%
%
More importantly, with the property of self-enforcement and public verification, smart contracts have become a major disruptive force to the legal sphere in the US, China, etc~\cite{us-smartcontract-court, yahoo-chinasupremecourt}.
\red{Swarm, the subproject responsible for Ethereum storage services, proposes the concept of storage auditing in the decentralized paradigm.}
\red{However, it falls short of an incentive layer and the fair auditing service is outsourced to reliable parties, given all the on-chain constraints of smart contract.}

As the leading force of decentralized storage system, Storj presents an auditing framework with centralized auditors called ``satellites''.
However, the selection of these auditors is subject to their reputation.
On this account, the entire system is built upon the chaining of reputation, where the reputation of each storage provider is individually determined by satellites. 
The problem is that the influential role of satellites may cause collusion and even amount to a de facto plutocracy, as the Storj founders and parties with close partnership take a leading role in maintaining and even controlling the auditing and payment transactions in the system. 
%
%

Siacoin~\cite{siacoin2014} is one of the pioneers that proposes a fully decentralized storage network with publicly verifiable file contracts on top of the blockchain consensus.
In its construction, storage providers prove the storage by periodically submitting part of the original file and the corresponding hashes within the file's Merkle tree to the blockchain.
However, the storage provider can reuse the proofs for challenged blocks to generate new proofs accordingly, instead of honestly storing all data faithfully, due to the low entropy of challenge randomness.

Along with this design, Filecoin~\cite{filecoin2017} presents an user-assisted auditing framework, with its focus on proof of replication for public data storage.
The submitted proof to the Filecoin blockchain reaches our requirement of on-chain perfect data privacy, as the proofs are produced with ZK-SNARK circuits, though the original intention of Filecoin is to leverage verifiable computation to compress the proofs.
However, the computational overhead during the trusted setup phase and the proof generation phase makes it hard to be deployed at present.

\noindent\textbf{Remarks.} It is noteworthy that despite the drawbacks we have pinpointed above, Storj and Siacoin are still under active research and plans of improvements.
At its early stage, Filecoin is still under heavy construction. 
We also consider two generic representatives of ZK-SNARK frameworks that protect the on-chain privacy, namely ZKCSP~\cite{zkcsp} for Bitcoin and Hawk~\cite{hawk} for Ethereum.

\section{Overview}
\label{sec-problem}
In this section, we aim to provide the background information, the system overview, and the adversarial model.

\subsection{Decentralized Storage Architecture}
\label{subsec-background}

\begin{figure}[!t]
  \centering
  \includegraphics[width=1.\linewidth]{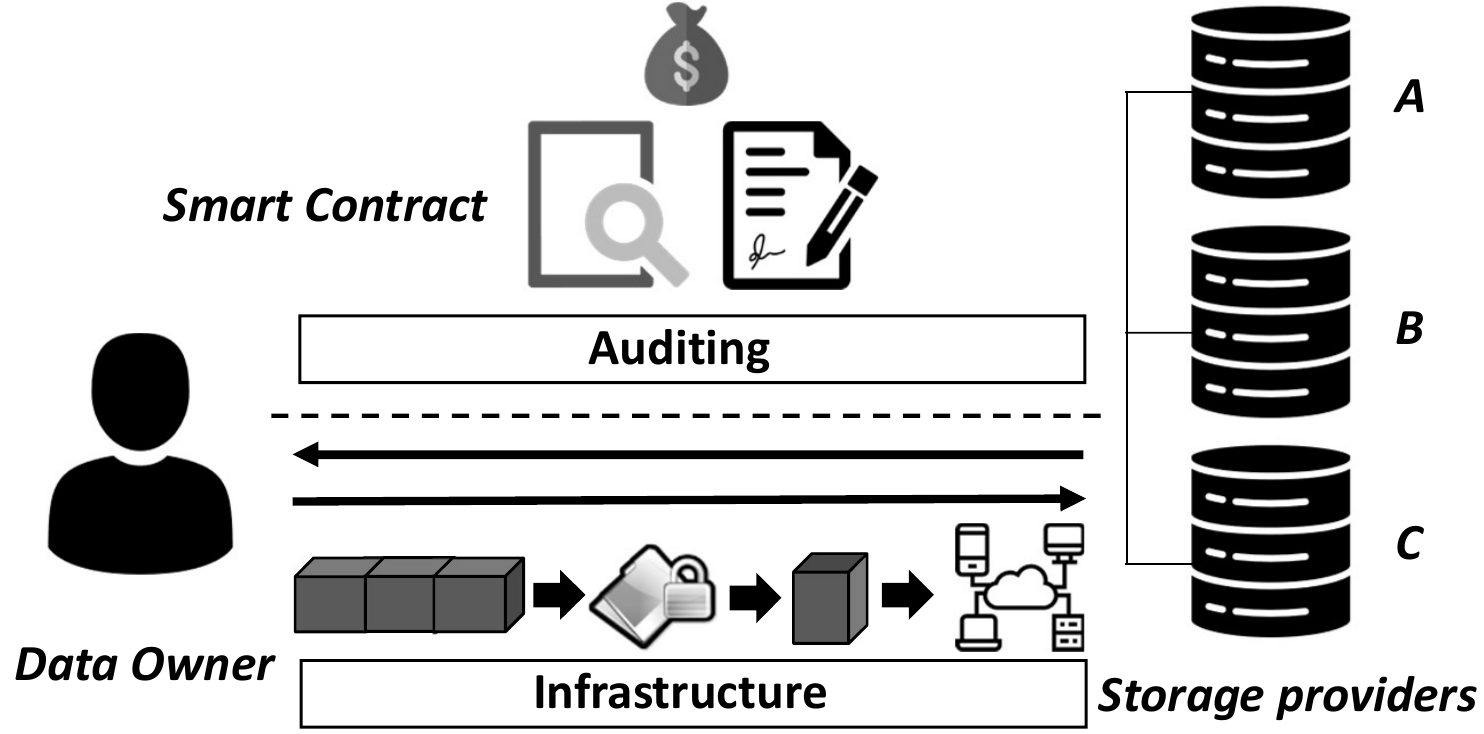}
  \caption{The architecture of the DSN data storage auditing.}
  \label{fig-flow}
  \vspace{-10pt}
\end{figure}

%
As illustrated in Fig.~\ref{fig-flow}, the system can be categorized into two parts, namely one for the storage infrastructure, and the other for auditing as well as incentives.
The storage infrastructure illustrated in Fig.~\ref{fig-flow}, is responsible for data chunking, data encryption, erasure coding, distributed networking, etc. 
Briefly, data to be outsourced is first chunked into pieces and encrypted at the block level by the data owner.
In addition, erasure coding~\cite{bowers2009proofs} (parity blocks) is also required for data redundancy.
At last, the data owner looks up the storage provider candidates using the distributed hash table~\cite{stoica2001chord} and uses this table for routing.

\noindent\textbf{Discussion on private-storage settings.} In our system, we apply data encryption and erasure coding for private archive data storage, and we stress that the encryption is a mandatory action taken on the side of the data owner.
To deal with public storage that cannot apply encryption, Filecoin leverages Data sealing~\cite{filecoin2017} and proof of replication~\cite{PoRep2018}, which is not required at all in our design.

\noindent\textbf{Further remarks.} During the following presentation, for simplicity, we only consider a one-to-one mapping between a data owner and a single storage provider.
That is, we assume a simplified scenario, where the data owner stores all data at one storage provider.
While for real-world overhead estimation, the data owners are free to adjust the data redundancy level and the number of storage providers, which leads to a linear increase on the total auditing cost.

\subsection{System Overview}
\label{subsec-overview}
We naturally focus on the design of auditing and incentives.
Specifically, there are three main roles in the framework, namely the data owner $\mathcal{D}$, who requests for the distributed storage of its encrypted data and later pays for each round of storage auditing;
the storage provider $\mathcal{S}$, who earns his deserved profits by storing the storage provider's files and cooperating with the storage auditing;
and the smart contract $\mathcal{SC}$ dedicated to the auditing job, who faithfully challenges the storage provider, validates the response proof in each round of the storage auditing protocol, and accordingly rewards the honest behaviours and punish the wrongdoers.

After the contract negotiation, $\mathcal{D}$ and $\mathcal{S}$ first go through a pre-processing phase and then the periodical auditing would carry on.
%
%
To achieve fairness in the blockchain model, both parties are required to submit deposits ahead of auditing, the smart contract is responsible for addressing all potential disputes between $\mathcal{D}$ and $\mathcal{S}$.
It further takes full control of the auditing and manages payments.

\noindent\textbf{Assumptions in the blockchain model.} In its essence, we view $\mathcal{D}$ and $\mathcal{S}$ as two rational players economically, who are fully incentivized to maximize their monetary rewards.
$\mathcal{SC}$ is modeled as an unbiased auditor, who bears the duty of auditing after $\mathcal{D}$ and $\mathcal{S}$ have an agreement upon the data to be outsourced, e.g., the contract duration, the monetary rewards, etc.
When the auditing begins, the smart contract extracts randomness from a secure randomness beacon~\cite{rabin1983beacon} and issues the challenge for each round of auditing. 
After receiving the response from $\mathcal{S}$, the smart contract enforces its pre-determined contract logic, thus maintaining the balance of the ecosystem in the decentralized world.
We stress that the auditing frequency should be set properly (at least every few hour), such that compared to the auditing interval, the time of block confirmation is much smaller.  

\begin{proto-func}{$\party{D}$ and $\party{S}$ have reached a consensus after contract negotiation}{Smart contract functionality for secure storage auditing. The contract is modeled as a state machine. We do not enumerate all functions in the real contract and roughly divide the procedures into the initialization phase and the auditing phase.}{fig:contract}
\begin{flushleft}
\violet{\textbf{Initialize:}}
\end{flushleft}
\begin{pevent}{On receive}{(\msg{negotiated}, \var{agrmts}, \var{params}, \var{metadata}) from $\party{D}$, $\party{S}$}
\pindent assert $\var{st} = \bot$;
\pindent set \var{agrmts};  \pcomment{e.g., \var{T} for contract duration, \var{num} for auditing times.}
\pindent set \var{params}, \var{metadata}; 
\pindent set $\var{st} \assign \const{ACK}$, $\var{cnt} \assign 0$;
\pindent broadcast $\msg{negotiated}$; 
\end{pevent}
\begin{pevent}{On receive}{(\msg{acked}) from $\party{S}$}
\pindent assert $\var{st} = \const{ACK}$; 
\pindent set $\var{st} \assign \const{FREEZE}$; broadcast $\msg{acked}$;
\end{pevent}
\begin{pevent}{On receive}{(\msg{freeze}, $\money{\mathcal{D}}, \money{\mathcal{S}}$) from $\party{D}$, $\party{S}$}
\pindent assert $\var{st} = \const{FREEZE}$;
\pindent lock $\money{\mathcal{D}}, \money{\mathcal{S}}$ for \var{T}; 
\pindent set $\var{st} \assign \const{AUDIT}$; broadcast $\msg{inited}$;
\pindent call scheduling(\msg{Chal}); \pcomment e.g., by Ethereum Alarm Clock service
\end{pevent}

\begin{flushleft}
\violet{\textbf{Audit:}}
\end{flushleft}
\begin{pevent}{On trigger scheduling}{$(\msg{Chal})$} 
\pindent assert $\var{st} = \const{AUDIT}$, \var{cnt} $\in \left[0, \var{num}\right]$;
\pindent output $R$ from Randomness Beacon;  
\pindent set $\var{st} \assign \const{PROVE}$; broadcast $\msg{challenged}$;
\end{pevent}

\begin{pevent}{On receive}{$(\msg{prove}, \var{prf})$ from \party{S}}
\pindent assert $\var{st} = \const{PROVE}$ and \var{cnt} $\in \left[0, \var{num}\right]$;
\pindent broadcast $\msg{proofposted}$; call scheduling(\msg{Verify});
\end{pevent}

\begin{pevent}{On trigger scheduling}{$(\msg{Verify})$} 
\pindent \pif $\var{cnt} \geq \var{num}$ : $\bot$; 
\pindent \pif $V(\var{params}, \var{metadata}, \var{prf})$ = True : 
\pindent[12] broadcast $\msg{pass}$; 
\pindent[12] unlock and transact $\money{}$ to $\party{S}$; 
\pindent \pelse : 
\pindent[12] broadcast $\msg{fail}$;
\pindent[12] unlock and transact $\money{}$ to $\party{D}$;
\pindent $\var{cnt}$++; call scheduling(\msg{Chal}); set $\var{st} \assign \const{AUDIT}$;
\end{pevent}
\end{proto-func}


\subsection{Adversarial Model}
\label{subsec-threat-adversarial}
%
First, we need to consider the fairness of incentives.
On the one hand, it is most natural for $\mathcal{S}$ to behave unfaithfully.
For instance, it may simply drop the data to reclaim more storage for more monetary benefits;
similarly, it may delicately discard data rarely accessed by $\mathcal{D}$;
it may also hide data loss incidents for the consideration of his or her own reputation.
On the other hand, an incentive-driven $\mathcal{D}$ also has motivation to behave unfaithfully.
In the extreme case, $\mathcal{D}$ may generate incorrect metadata for the storage provider such that the auditing would always favor the data owner's financial benefits.
Assuming rational economic actors, we rule out the possibility of malicious behaviors by $\mathcal{D}$ and $\mathcal{S}$ if they are faced with financial penalty.
Later in Section~\ref{completeNsound}, though, we discuss possible countermeasures to defend against such attacks.
%

We also need to consider \red{adversarial behavior targeting on-chain audit trails}. We treat the blockchain as a trusted party under the honest majority assumption, in the sense that it stores the audit trails in a fully transparent manner. 
Due to the transparency and tamper-proof property of the blockchain, an off-chain adversary can \red{observe on-chain audit trails stealthily and extract knowledge of raw data unknowingly.} 
%
%
\red{We emphasize that in spite of being a seemingly small security loophole, such vulnerability could be easily leveraged and also augmented by adversaries to launch more sophisticated attacks. For more details, see our full analysis in~\ref{subsec-design-rand}.}
%

\section{Strawman proposal}
\label{sec-strawman}
To present our main protocol smoothly, we give the following strawman proposal that focuses on the seamlessly integration between auditing and the achievement of fair payments.
Straightforwardly, we leverage a standard set of primitives that offer on-chain privacy and efficiency simultaneously as the strawman auditing solution.
%

\subsection{Primitives}
Generally, zero knowledge proofs can convince the verifier of the validity of a particular statement without leaking any other information beyond the validity of the statement itself.
Recent progress on practical and generic zero knowledge proof systems~\cite{parno2013pinocchio,ben2013snarks4C} makes them become widely adopted tools to achieve proof succinctness and quick verification on the blockchain.
It is noteworthy that a number of tools on zero knowledge proving systems are already available to support arbitrary computation by implementing general-purpose compilers from high-level languages.
Among all the zero knowledge proving systems, ZK-SNARK~\cite{ben2014succinct,groth2016size} is by far the most useful construction.
\vspace{-5pt}

\subsection{Basic Instantiation}
\label{subsec-contract}
As a basic instantiation of the auditing protocol, we can encapsulate a Merkle-tree based auditing protocol within a pre-built ZK-SNARK circuit to achieve on-chain privacy and on-chain efficiency.
We first assume the trusted setup has been accomplished for the ZK circuits and it outputs the proving key $pk$ and the verification key $vk$.
Also, before auditing, $\mathcal{D}$ needs to construct a Merkle tree from data to be stored and obtain the Merkle root $rt$, which can be publicly accessible.
After $\mathcal{S}$ receives the data from $\mathcal{D}$, the auditing starts as forms of challenge-response protocols.

During the contract duration, for each challenge $R$ issued, $\mathcal{D}$ finds the challenged block $m_{i}$ (leaf node in the Merkle tree) with a pseudo-random function.
To prevent public verification from leaking data information, $\mathcal{D}$ leverages the established ZK-SNARK circuit and $pk$ to generate a zero knowledge proof $prf$.
The proof validates the statement that the challenged leaf node $m_{i}$ and the corresponding Merkle path $path$ can always lead to $rt$ without leaking any information beyond the statement.
With $vk$ and challenge $R$, the produced $prf$ can be verified efficiently with on-chain privacy.

\subsection{Disputes and fairness}
Disputes could occur even before auditing, which is the $Initialize$ phase in Fig.~\ref{fig:contract}.
After $\mathcal{D}$ has deployed a contract with contract details $argmts$, $params$ containing $pk, vk$, and $metadata$ in the form of $rt$, $\mathcal{S}$ has to agree on them before the contract can carry on.
Notice that $\mathcal{D}$ cannot gain profits by forging $params$ and $metadata$, as $params$ and $metadata$ are generated during the trust setup of the ZK-SNARK circuit for the Merkle tree construction over the data to be stored.
Therefore, in most cases, $\mathcal{S}$ simply sends an acknowledgment signal to the smart contract and then the smart contract awaits both parties to deposit cryptocurrencies.
While in some extreme circumstances, $\mathcal{S}$ may act maliciously, terminating the contract and thus making $\mathcal{D}$ pay the on-chain storage fees for $argmts$, $params$, and $metadata$.
In Section~\ref{completeNsound}, we briefly discuss possible countermeasure.

When the periodical $Audit$ phase in Fig.~\ref{fig:contract} starts immediately after the deposits, it is easier to achieve fairness.
If $\mathcal{S}$ submits $prf$ that can be verified by $vk$, $\mathcal{S}$ would obtain micro-payments from the deposits locked by $\mathcal{SC}$;
In contrast, if $prf$ cannot be verified by $vk$, $\mathcal{D}$ would obtain micro-payments.
The above procedures continue with pre-determined auditing frequency, until eventually the contract expires.

\subsection{Limitations}
\label{subsec-limit}
The strawman solution that uses generic zero knowledge proofs is notoriously hard to be deployed and also far from optimal, as a tremendous amount of overhead is forced during the off-chain procedures, including the trusted setup phase and the proof generation phase.
%
%
Besides, challenge randomness would be exhausted and after that attacks can simply exploit the reused challenge.  
In light of the limitations of generic ZK tools, we need an alternative auditing scheme.

\section{Our Main Protocol}
\label{sec-design}
With a streamlined smart contract template and a basic instantiation explained, we now introduce the design rationale of our main protocol and present our protocol details.

\subsection{Design Rationale}
As we have \red{mentioned}, it creates a tremendous amount of off-chain overhead by straightforwardly employing the generic zero knowledge proving systems.
To design a pragmatic auditing protocol with on-chain privacy, on-chain efficiency, and on that basis also with the consideration of off-chain efficiency, we turn to the public-key cryptography, more specifically the homomorphic linear authenticators (HLA).
\red{See more related work in Section~\ref{relatedwork}.}

Basically, HLA encapsulate public homomorphic authenticators for public verification.
Extensive research on cloud storage auditing~\cite{wang2009ensuring,wang2013privacy} proves its theoretical asymptotic efficiency.
However, this line of work has yet to be deployed on a large scale in practice due to the efficiency consideration.
Indeed, most designs would bring quite an amount of processing time for pre-processing and extra storage upon the storage provider.
Though this overhead can be alleviated, as suggested in ~\cite{shacham2008compact}, it further leads to the increase of proof size.
In order to improve efficiency, the polynomial commitment can be leveraged to bring down the storage overhead and processing time without raising the proof size~\cite{xu2012towardsefficientpor,yuan2013proofs}.

Without the use of generic ZK tools, applying the efficient polynomial commitment upon HLA could be dangerous in the paradigm of decentralized storage. 
In our following presentation, \red{we point out the real-world impact of straightforwardly applying a non-ZK version of HLA-based auditing protocol.}
\red{We show that potential adversaries could leverage the ``extractable knowledge'' sealed in on-chain proofs, and accelerate the attacking process with other possible attacks in the real-world scenarios of the blockchain paradigm.}
On this account, by leveraging the Sigma protocol~\cite{schnorr91} that fully utilizes the algebraic structures of HLA to hide the data, we propose our synergistic design to achieve on-chain privacy, on-chain efficiency, and off-chain efficiency at the same time.
%
%

\subsection{Audit Details}

\begin{figure*}[t]
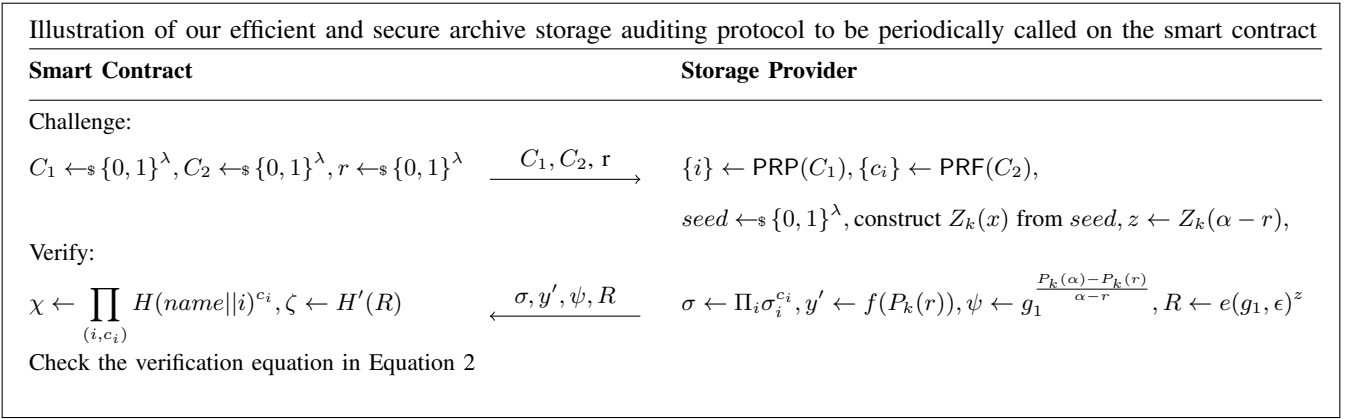

\centering
\fbox{
\procedure{Illustration of our efficient and secure archive storage auditing protocol to be periodically called on the smart contract}{
\textbf{Smart Contract} \< \< \< \< \< \<  \textbf{Storage Provider} \\ 
\pchline 
\pccolone{\text{Challenge:}} \\
\pccolone{C_1 \sample \bin^{\lambda}, C_2 \sample \bin^{\lambda}, r \sample \bin^{\lambda}} 
\pccoltwo{\sendmessageright{width=1cm, length=2cm, top=\text{$C_1, C_2$, r}}} 
\pccolthr{\{i\} \gets \prp(C_{1}), \{c_{i}\} \gets \pr(C_{2}), } \\ 
\< \< \< \< \< \< seed \sample \bin^{\lambda}, \text{construct} \; Z_{k}(x) \; \text{from} \; seed, z \gets Z_{k}(\alpha - r), \\
\pccolone{\text{Verify:}}\\
\pccolone{\chi \gets \prod_{(i,c_{i})}  H(name||i)^{c_{i}}, \zeta \gets H'(R)} 
\pccoltwo{\sendmessageleft{width=1cm, length=2cm, top=\text{$\sigma,  y', \psi, R $}}} 
\pccolthr{\sigma \gets \Pi_{i} \sigma_{i}^{c_{i}}, y' \gets f(P_{k}(r)), \psi \gets g_{1}^{\frac{P_{k}(\alpha)-P_{k}(r)}{\alpha-r}}, R \gets e(g_{1},\epsilon)^{z}} \\
\pccolone{\text{Check the verification equation in Equation~\ref{eq:verify}}} \\
}
}
\caption{Flowchart of the auditing interaction during our enhanced auditing protocol}
\label{auditing-proto}
\vspace{-10pt}
\end{figure*}

Figure~\ref{auditing-proto} displays the interactions between the smart contract and the storage provider. 
For simplicity, we omit the contract template and detailed procedures described in Section~\ref{subsec-contract}.
Denote the security parameter as $\lambda$.
We use an asymmetric bilinear map $e: \mathbb{G}_{1} \times \mathbb{G}_{2} \rightarrow \mathbb{G}_{T}$ in our following auditing protocol, where $\mathbb{G}_{1}$ and $\mathbb{G}_{2}$ are multiplicative cyclic groups. And $\mathbb{G}_{1}$ has prime order $p$.
Let a group generator $g_{1}$ selected randomly from $\mathbb{G}_{1}$, a group generator $g_{2}$ selected randomly from $\mathbb{G}_{2}$ and specify a random oracle $H$ : $\{0,\ 1\}^{*} \rightarrow \mathbb{G}_{1}$. 
Assume the file to be stored as $F$.
It is further divided into $n$ data blocks in the form of group elements.
Then, each $s$ collections of data blocks can constitute data chunks for the acceleration of data processing and the savings of extra storage overhead.
Formally, $F$ is equivalent to the collection of chunks \{${m}_{i} = (m_{i,0}, \cdots, m_{i,s-1}) \in (\mathbb{Z}_{p})^{s}\}_{i = 0}^{\ceil{\frac{n}{s}}}$, where we use \ceil{x} to indicate the minimal integer larger than or equal to $x$.
Note that the last data block may need padding.
\begin{defn}
Denote $d = \ceil{\frac{n}{s}}$; $M_{i}(x) =  m_{0, i}+m_{1, i}\cdot x^{1}+\cdots+ m_{s-1, i}\cdot x^{s-1} \bmod p$ for $i \in [0,d)$ is a polynomial commitment of data blocks with the degree being $s-1$; 
\end{defn}
\begin{defn}
Pseudo Random Permutation function $\pi: \bin^{\lambda} \times \bin^{\log{n}} \rightarrow \bin^{k}, (k<\lambda)$ is a function that cannot be distinguished from random permutation;\\
Pseudo Random Function $f: \bin^{\lambda} \rightarrow \mathbb{Z}_{p}^{k}, (k<\lambda)$ is a function that cannot be distinguished from random function.
\end{defn}
\noindent\textbf{Initialize.} During the generation phase of public parameters, $\mathcal{D}$ should randomly sample two group elements $\alpha, x \sample \mathbb{Z}_{p}$, and set them as the private key.
Meanwhile, compute $\epsilon \leftarrow g_{2}^{x}$, $\delta \leftarrow g_{2}^{\alpha x}$ and $\{ {g_{1}}^{{\alpha}^{j}} \}_{j=0}^{s-2}$.
The public key $pk$ is $(p, \epsilon, \delta,\{ {g_{1}}^{{\alpha}^{j}} \}_{j=0}^{s-2}, g_{2}, e(g_{1}, \epsilon), H)$ and the private key $sk$ is $(x,\alpha)$.
To generate metadata, i.e., homomorphic authenticator, we utilize the pairing-based polynomial commitment~\cite{KateZG10,libert2016functional} of the group elements of each block as illustrated below. 
%
%
Making use of this polynomial, $\mathcal{D}$ binds all groups inside a block to a authenticator.
%
%
To generate the homomorphic authenticators, $\mathcal{D}$ also needs to sample a file identifier called $name$ from $\mathbb{Z}_{p}$ such that $H(name||i)$ can be used for block indexing.
Noticeably, $name$ is also recorded on the blockchain for public verification.
Putting it together, the authenticators of each chunk thus becomes $\sigma_{i} = (g_{1}^{M_{i}(\alpha)} \cdot {H(name||i)})^{x} \in \mathbb{G}_{1}$.

After $\mathcal{D}$ transfers data to be stored and its authenticators in a secure channel, $\mathcal{S}$ checks it with public keys and sends a signal to the smart contract to either carry on or terminate the contract.
Note that the chance of $\mathcal{D}$ forging authenticators is negligible after this initial check.

\noindent\textbf{Challenge.} Due to the throughput and cost-effectiveness of randomness produced on the blockchain, it is more wise for $\mathcal{S}$ to use pre-determined algorithms to expand the domain of randomness outputs.
Specifically, upon receiving the random challenges consisting of \{$C = (C_1, C_2), r$\} issued from the smart contract, $\mathcal{S}$ needs to use $\pi$ and $f_{1}: f$ to derive $\{i\}_{i=0}^{k-1}$ and $\{c_{i}\}_{i=0}^{k-1}$, where $k$ is the number of challenged chunks.

\begin{defn}
Polynomials derived from or related to $M_{d}(x)$: \\
$P_{k}(x) = \sum_{j=0}^{k-1} c_{j}m_{j, 0}+\sum_{j=0}^{k-1} c_{j}m_{j, 1}\cdot x^{1}+\cdots+\sum_{j=0}^{k-1} c_{j}m_{j, s-1}\cdot x^{s-2} \bmod p$ is a polynomial commitment of linear combinations of $k$ challenged chunks;\\
The polynomial quotient $Q_{k}(x) = \frac{P_{k}(x)-P_{k}(r)}{x-r} \equiv \beta_{0}+\beta_{1}\cdot x^{1}+\cdots+\beta_{s-1}\cdot x^{s-1} \bmod p$;
And we have $g_{1}^{Q_{k}(\alpha)} = \prod_{i=0}^{s-1} (g_{1}^{{\alpha}^{i}})^{\beta_{i}}$ \\
\vspace{-12pt}
\end{defn}

\noindent\textbf{Audit.} Briefly, $\mathcal{S}$ generates a proof by computing the authenticator $\sigma$, the polynomial commitment $P_{k}(r)$, and $\psi = g_{1}^{Q_{k}(\alpha)}$ from public keys as well as polynomial quotient coefficient. 

For proof verification, by leveraging the techniques introduced in~\cite{yuan2013proofs}, we can derive the following equation:
\begin{equation}
e(\sigma, g_{2})\cdot e(g_{1}^{-P_{k}(r)}, \epsilon) = e(\chi, \epsilon)\cdot e(\psi, \delta\cdot\epsilon^{-r})
\end{equation} 
In the above equation, $\chi$ is obtained by computing $\chi = \prod_{(i,c_{i})}  H(name||i)^{c_{i}}$, denoted as $\prod_{i} t_{i}$.

\subsection{Impact of On-chain Privacy}
\label{subsec-design-rand}
\red{One security problem to worry about is that} malicious adversaries are able to extract data contents from the polynomial commitment within the audit trails.
To see why, suppose an adversary observes a sufficiently long sequence of challenges and proofs.
Denote the sequence contains $s\cdot u$ pairs of challenges $(r_{i, j}, C)$ and proofs $(\sigma_{i, j}, \{P_{k}(r)\}_{i, j}, \psi_{i, j})$, where $i \in [0, s), j\in [0, u)$.

Since a polynomial with degree $s-1$ can eventually recovered through the Lagrange polynomial interpolation, given $s$ points on the polynomials, the adversary is thus able to reconstruct $P_{k}(x)$ using the observed challenges and proofs.
At this stage, the adversary already recovers $\{c_{j}\cdot m_{j}\}_{(0\leq j \le u)}$ from polynomial coefficients.
According to the analysis in ~\cite{shacham2008compact,wang2013privacy}, the adversary can easily extract data knowledge from this linear combination of data blocks given $u$ unknown data blocks plus $u$ pairs of challenges and proofs.

Indeed in our usage scenario, it creates an opportunity for anonymous adversaries to exploit the above attacks and recover partial data, by simply observing and computing from accumulated challenges and proofs on the blockchain.
The situation gets even worse when the data owner selects a polynomial with a small degree.
In the extreme case of data of small size (the number of blocks is roughly the same as the challenge domain), every single block can be recovered by adversaries given a normal contract duration (e.g., years).

Apart from the direct exploitation of on-chain audit trails, the adversaries may leverage more advanced attacks to recover the original data of $\mathcal{D}$.
For instance, adversaries can launch eclipse attacks~\cite{eclipse-bitcoin} with low resources~\cite{eclipse-ethereum} to isolate the victims from the rest of peers in the public blockchain network.
After monopolizing the victims' view of the blockchain, the adversaries can then issue a set of well-calculated challenge randomness and on that basis, extract the original data using the above-mentioned method much more efficiently.

\subsection{Secure Audit}
\label{subsec-secure-audit}
In view of this, we provide our solution to achieve the on-chain privacy guarantees.
For the following presentation, we start from the storage provider, who needs to respond to the challenge by submitting a proof.

\textit{1) Computation of auditing proofs.} $\mathcal{S}$ can compute the proof using the expanded randomness and polynomials defined above.
First, $\mathcal{S}$ produces the aggregated authenticator of the challenged chunks $\sigma = \Pi_{i} \sigma_{i}^{c_{i}} \in \mathbb{G}$ for $i \in [0, k)$.
Then, $\mathcal{S}$ computes $\psi = g_{1}^{Q_{k}(\alpha)}$ from the public key$\{ {g_{1}}^{{\alpha}^{j}} \}_{j=0}^{s-1}$ and coefficients calculated using the finite field polynomial quotient algorithm in $\mathbb{Z}_{p}$, as $\alpha$ is unknown to $\mathcal{S}$.
Instead of computing $P_{k}(r)$ that leaks data contents, $\mathcal{S}$ needs to randomly generate a hiding parameter $z \sample \mathbb{Z}_{p}$ and use it to further generate another hiding parameter $\zeta = H'(R)$, where $R = e(g_{1}, \epsilon)^{z}$ and the random oracle $H': \mathbb{G}_{T} \rightarrow \mathbb{Z}_{p}$ is universal and pre-determined.
Denote $y = P_{k}(r)$, then $\mathcal{S}$ can obtain $y' = \zeta \cdot P_{k}(r) + z$, as forms of Sigma protocols.
The final response from $\mathcal{S}$ would be $(\sigma, y', \psi, R)$, which is recorded on the blockchain along with the challenge randomness.

\noindent\textbf{Remarks.} To apply the Sigma protocol in the strict sense, each coefficient of the polynomial $P_{k}(x)$ requires to be affine masked with hiding parameters in the finite field and we denote the new polynomial $P'_{k}(x)$.
Note in practice, though, the hiding parameters $z$ and $\zeta$ is statistically indistinguishable from $P'_{k}(x)$ for any probabilistic polynomial time adversary.
%
%
%
%
%
%

\textit{2) Procedures of on-chain verification.} With the response posted, the smart contract automatically triggers the verification of the storage proof and runs the embedded algorithm logic.
%
%
%
During the verification, the smart contract first derives $\chi = \prod_{i} t_{i}$ from the public parameter $name$ and the two seeds.
Next, the smart contract also computes $\zeta \leftarrow H'(R)$.
To validate the proof, the smart contract checks the equation: 
\begin{equation}
R \cdot e(\sigma^{\zeta}, g_{2}) \cdot  e(g_{1}^{-y'}, \epsilon) = e(\chi^{\zeta},\epsilon)\cdot e(\psi^{\zeta},\delta\cdot\epsilon^{-r})
\label{eq:verify}
\end{equation}
It is straightforward to see the above equation holds:
\begin{equation*}
\begin{aligned}
LHS &= R \cdot e((\prod_{i} t_{i})^{x\zeta} \cdot g_{1}^{x\zeta\cdot P_{k}(\alpha)}, g_{2}) \cdot e(g_{1}^{-(\zeta \cdot P_{k}(r)+z)}, g_{2}^{x}) \\
& = R \cdot e(g_{1}, g_{2})^{x(\zeta \cdot (P_{k}(\alpha)-P_{k}(r))-z)}\cdot e(\prod_{i} t_{i}, g_{2})^{x\zeta}  \\
& = e(g_{1}, g_{2})^{x\zeta \psi (\alpha - r)}\cdot e(\prod_{i} t_{i}, g_{2})^{x\zeta} = RHS
\end{aligned}
\label{eq:deduction}
\end{equation*}

\subsection{Reliable challenging randomness}
\label{subsec-design-rand}
One critical issue that affects the correctness and privacy of proofs is the generation of reliable, unpredictable, unbiased randomness.
Previously, we assume the smart contract can simply request from the randomness beacon.
But how can we efficiently obtain adequate randomness in practice?
The most straightforward solution is to call on a number of data owners to play commit-and-reveal games~\cite{syta2017RandHerd,schindler2018hydrand}, which is the core idea behind current implementation of Randao~\cite{randao} and many similar services on Ethereum. 
However, in recent analysis~\cite{chatterjee2019probabilistic}, it shows that the last participant to submit partial randomness may have a way of maneuvering its partial to favor its own interest.
To this end, recent work~\cite{bunz2017proofs} uses the concept of verifiable delay function to fix this loophole.
Alternatively, we can also introduce the extra assumption of a trusted party, e.g., temporal blockchain from NIST quantum randomness beacon~\cite{temtum}, and directly absorbing randomness from these trusted sources. 
In Section~\ref{subsec-onchain}, we show the cost of randomness generation on the blockchain.

\section{Security Analysis}
\label{sec-security}

In this section, we briefly evaluate the security of our secure auditing protocol of distributed archive storage enforced by the smart contract by analyzing the fulfillment of the security guarantees listed in Section ~\ref{subsec-threat-adversarial}.

\subsection{Storage correctness and fairness}
\label{completeNsound}
We have elaborated the completeness of our secure storage auditing protocol, i.e., a correct proof would always pass the verification, as shown in Sec.~\ref{eq:deduction}.
Hence, the interests of $\mathcal{S}$ can always be protected during the auditing period.

To show that a malicious $\mathcal{S}$ cannot forge a proof detrimental to the interests of $\mathcal{D}$, we first give a proof sketch of the extractability of auditing proofs submitted by an honest $\mathcal{S}$.
Essentially, the unforgeable problem can be transformed into the extractability of knowledge in a proof of knowledge problem.
More specifically, given two valid proof responses $(\sigma_{1}, y'_{1}, \psi_{1}, R_{1})$ and $(\sigma_{2}, y'_{2}, \psi_{2}, R_{2})$, unless the adversary can break CDH and q-BSDH assumption~\cite{boneh2004short}, it is feasible to extract partial linear combination of data blocks with the overwhelming probability.
Our analysis in the main protocol has shown some insights.
For more proof details, we refer the readers to the results in~\cite{shacham2008compact,yuan2013proofs,improvedfuzzy}.

Aside from the data extractability assured in theory, it is still crucial to specify $k$ to guarantee a high confidence level that the data stored is not tampered.
Plenty of previous studies have analyzed the relationship between the number of challenged chunks $k$ and the storage confidence level~\cite{ateniese2007provable}.
Particularly, setting $k$ to $300$ can give $\mathcal{D}$ storage assurance of $95\%$ if only $1\%$ of entire data is tampered. 
In the paradigm of decentralized archive storage, we believe this amount of challenged chunks is adequate to protect the interests of $\mathcal{D}$.

Given the above analysis of extractability and practical confidence of storage assurance, we have the following theorem:
\begin{thm}
 The storage provider cannot generate a proof to make the verification equation hold when he does not keep the file intact as it is and the number of challenged chunks is large enough in the random oracle, given the q-BSDH assumption.
\end{thm}

\noindent\textbf{Remarks on fairness in practice.} In our threat model, $\mathcal{D}$ and $\mathcal{S}$ rational players.
While in real-world deployments, a party may still execute the auditing protocol in a way that harms others' interest.
For instance, in the initialization phase, $\mathcal{S}$ can always send the rejecting signal to the smart contract and let $\mathcal{S}$ pay on-chain storage cost of public keys.
We stress this kind of denial-of-service attack would be good to none but worse to himself under a robust reputation-based system.
Using similar countermeasures, other attacks such as the Sybil attack~\cite{douceur2002sybil}, can also be alleviated.

\begin{table*}[!t]
\centering
\normalsize
\begin{threeparttable}
\setlength{\tabcolsep}{0.45 em}
\captionsetup{justification=centering}
\caption{Comparison of SNARK-based strawman solution and our main solution.}
\label{tab-exp}
\begin{tabular}{c|c|ccc|ccc|c}
&File Info.  & \multicolumn{3}{c}{Pre-process$^\dag$  }      & \multicolumn{3}{c}{Proof Generation}       &Verification              \\ \toprule
&Size    & Time                           & Param. size       & \# Constraints                     & Time                               & Memory            & Size     & Time      \\ \toprule
Strawman solution$^{\ast}$   &1 KB   & 260 s                              & 150 MB               & $3 \times 10^5$       & 30 s                               & $\sim$ 300 MB               & \textbf{384 bytes} & \textbf{30 ms}  \\
Our main solution   &1 GB    &$\sim$ 120 s                                   & $\sim$ 5 KB                    & -                              & 46 ms                              & 3 MB                 & \textbf{288 bytes} & \textbf{7 ms}      \\ \bottomrule
\end{tabular}
\begin{tablenotes}
\small
\item Note that all our evaluation is carried out with quad-core CPUs.
Also, our main solution supports data of much larger size and assures stronger storage guarantees, as in the Merkle tree auditing, only one Merkle path is examined in our experiment.
\item $\ast$ We leverage the Rust Bellman ZK-SNARK library~\cite{bellman} to implement a proof-of-concept Merkle tree based auditing prototype.
We stress the maximum data size allowed in current implementation is around 16 KB, as suggested in the work~\cite{xie2019libra}.
\item $\dag$ For the pre-preprocessing phase of our strawman solution, the trusted setup is required for ZK-SNARK circuit during the pre-processing phase.
The number of constraints required is closely related to the circuit size.
Public parameters encapsulate the proving keys, verification keys, etc.
While for the pre-processing phase of our HLA-based main solution, the data owner requires to generate the authenticators and the storage providers can validate the authenticators.
Note there is no circuit constraints involved.
\end{tablenotes}
\end{threeparttable}
\end{table*}

\subsection{Data privacy on the blockchain}
In this part, we show our secure auditing protocol can prevent data leakage from the auditing proofs stored on the blockchain.
Abstractly, as suggested in~\cite{rsa-pos}, it can be seen as a witness-indistinguishable Sigma protocol~\cite{cramer1996modular} for the relation:
$$\mathcal{R}= \left\{(pk, \{t_{i}\}, chal, \sigma), F \  \middle| \   \displaystyle \sigma^{pk}\equiv g_{1}^{F}\cdot \prod_{i}t_{i} \bmod N \right\},$$
$\tilde{F}$ is the witness in the above relation that counts as private inputs, which is essentially a linear combination of $F$.

As a relation with the property of witness indistinguishability, we can transform it into a more generic relation with stronger security property of zero knowledge.
In fact, we can further view the authenticator $\sigma$ as private inputs and prevent it from being brutal-force attacked by off-chain adversaries.
However, it would not be necessary to hide $\sigma$ in practice.
We stress that the theorem below can be proven:
\begin{thm}
For any probabilistic polynomial time adversary $\mathcal{A}_{PPT}$, the probability $\mathcal{A}_{PPT}$ can distinguish witness $(w_{1}, w_{2})$ after seeing all public inputs is negligible.
Moreover, under the standard Discrete Logarithm assumption, the probability $\mathcal{A}_{PPT}$ can distinguish $F$ from random data after seeing all public inputs are negligible.
\end{thm}
Briefly, for the case of distributed archive storage, the property of witness indistinguishability is adequate to achieve perfect data privacy on the blockchain.
Due to the space limit, we will not present a formal analysis.
Yet the underlying reason is that the encrypted data domain, along with its generated authenticators provides plenty of entrophy to prevent brutal-force attacks even if the adversaries have access to authenticators stored on the blockchain.

\section{Evaluation}
\label{sec-eval}
We now evaluate our auditing protocol by answering the following questions.
%
%
First and foremost, what is the on-chain overhead of the introduced operations?
On the basis of that, what is the overall compulsory auditing cost? 
Second, what is the off-chain overhead of our DSN auditing system?
Does it cause heavy usage cost for the data owners and the storage providers in practical usage?
Moreover, what is the scale of user base the system can support? 

\subsection{Implementation and experiments setup}
\label{subsec-setup}
Our implementation follows the design principle of compatibility and could be deployed as a plug-in component suitable for most underlying P2P-akin storage system and incentive systems constructed on the blockchain.
In particular, we choose Tahoe-LAFS~\cite{tahoe} and Ethereum as our testbed.
The major challenge we encounter on the Ethereum platform is that, it is impossible to efficiently implement most complicated cryptographic primitives natively due to the restrictions of the on-chain programming language Solidity.
\red{This obstacle hinders traditional testing approaches of Ethereum Testnet.}
\red{Though solid progress of more native language like WebAssembly~\cite{ewasm} has been proposed in the Ethereum community, we find it still requires heavy workload of development in the foreseeable future.}

\red{In view of this, we develop our pre-compiled contract with \textbf{opcode optimization}.}
Our contract, along with the off-line implementations, is consisted of over 1,000 \red{lines of code}, mostly in Golang.
%
%
Based on our customization of opcode, we deploy our own private testnet \red{with our preliminary proof-of-concept implementation}.
\red{We emphasize our testing approach is compatible with the current development plan of the WebAssembly framework proposed by the Ethereum community.}
\red{To simulate the basic storage services in the decentralized archive storage network, the private network is set up consisting of} three nodes, namely one representing the miner, one for the storage provider, and the other for the data owner.
%
%
The miner and the storage provider use Dell Poweredge T140 servers (Intel Xeon E-2174G CPU x4 @ 3.80 GHz) in Linux (Ubuntu Server 18.04 LTS).
The data owner uses a Desktop PC (Intel Core i7 8700k CPU x6 @ 3.70 GHz).
%
%

\begin{figure*}[t]
\begin{minipage}{0.33\linewidth}
\centering
\includegraphics[width=1.0\linewidth]{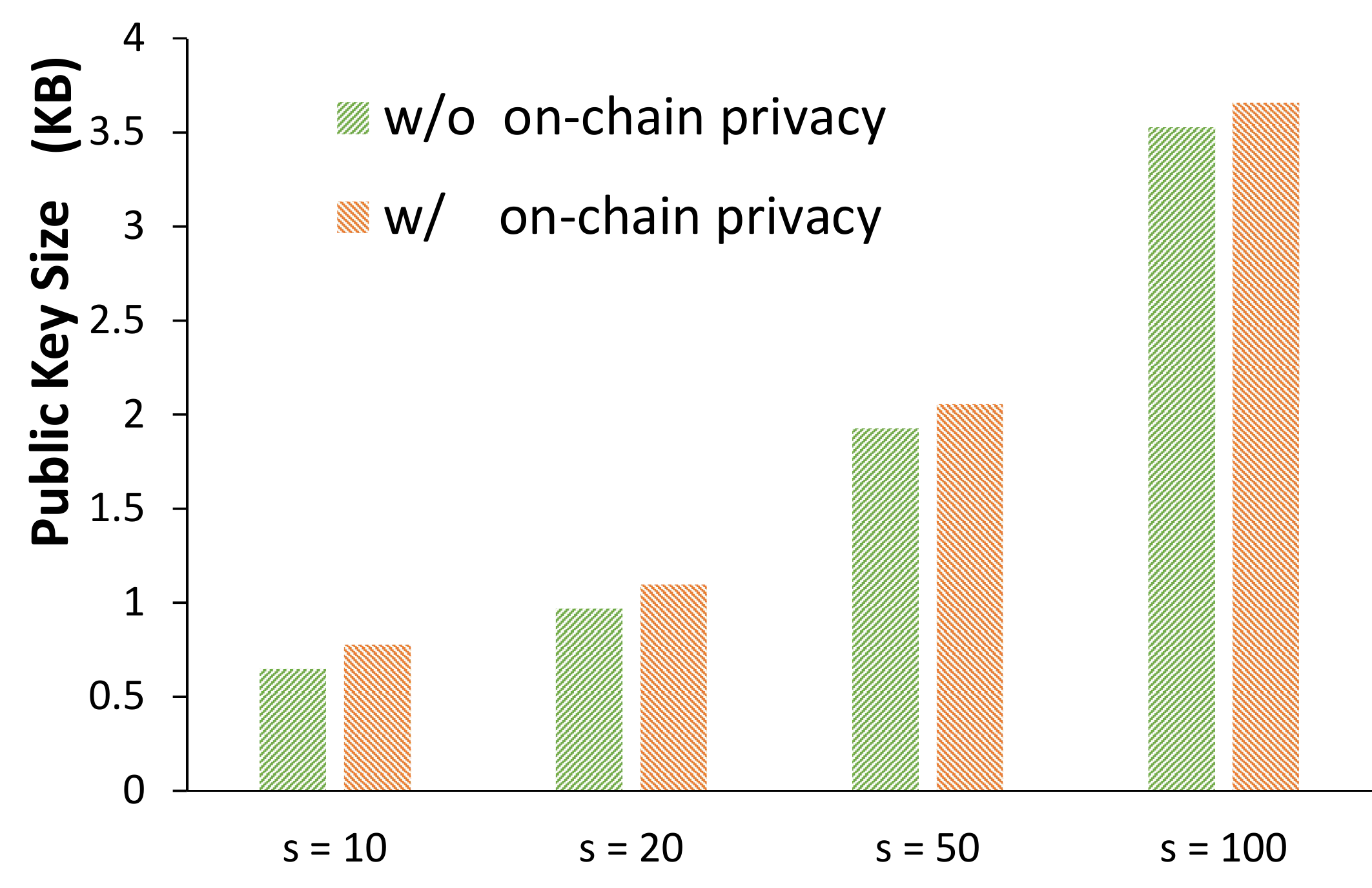}
\captionsetup{width=0.95\textwidth}
\caption{The initial one-time on-chain cost for public keys to be recorded.}
\label{fig:keysize}
\end{minipage}
\begin{minipage}{0.33\linewidth}
\centering
\includegraphics[width=1.0\linewidth]{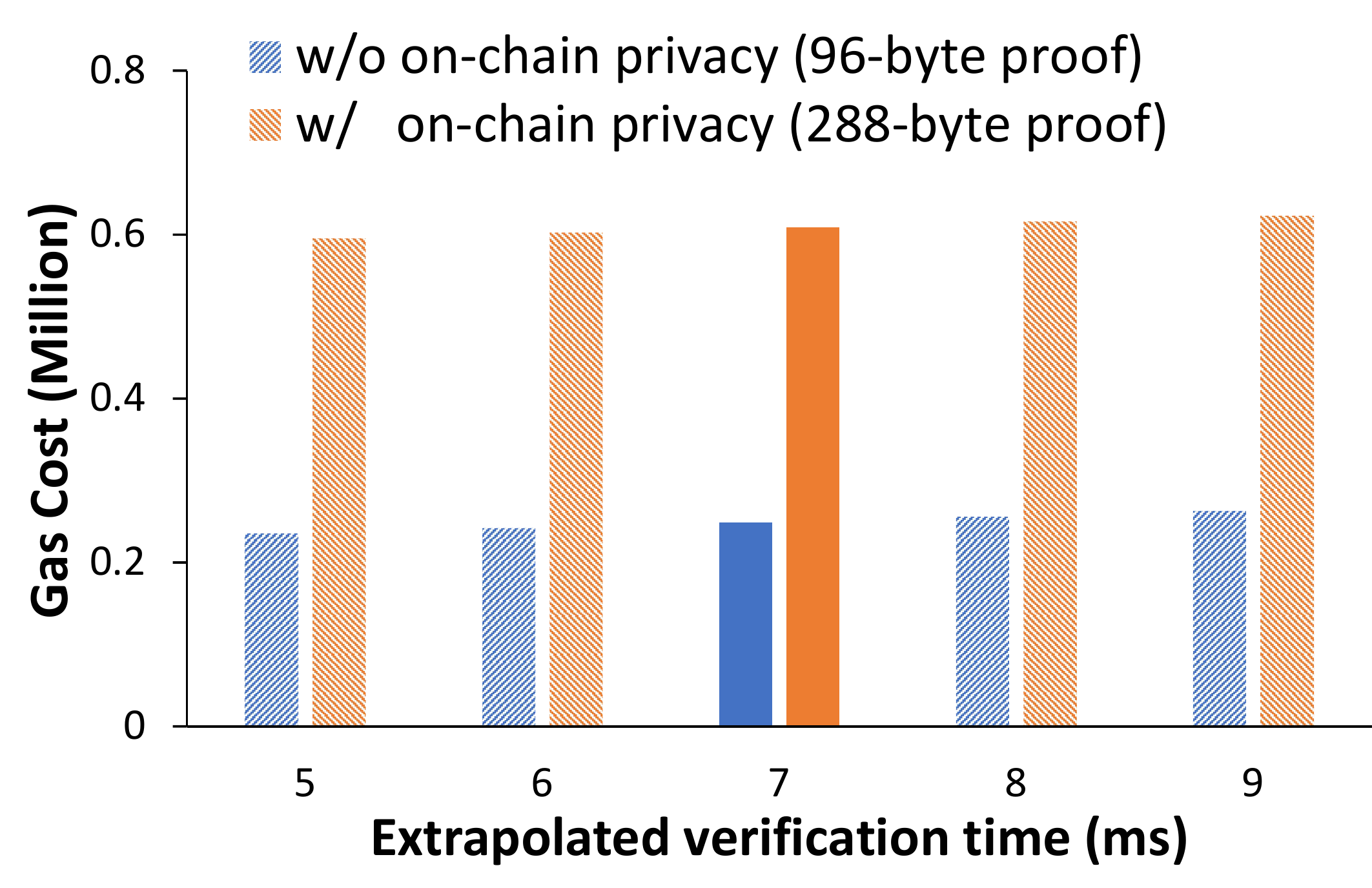}
\captionsetup{width=0.95\textwidth}
\caption{Gas cost calculation from proof size and extrapolated verification time.}
\label{fig:verify}
\end{minipage}
\begin{minipage}{0.33\linewidth}
\centering
\includegraphics[width=1.0\linewidth]{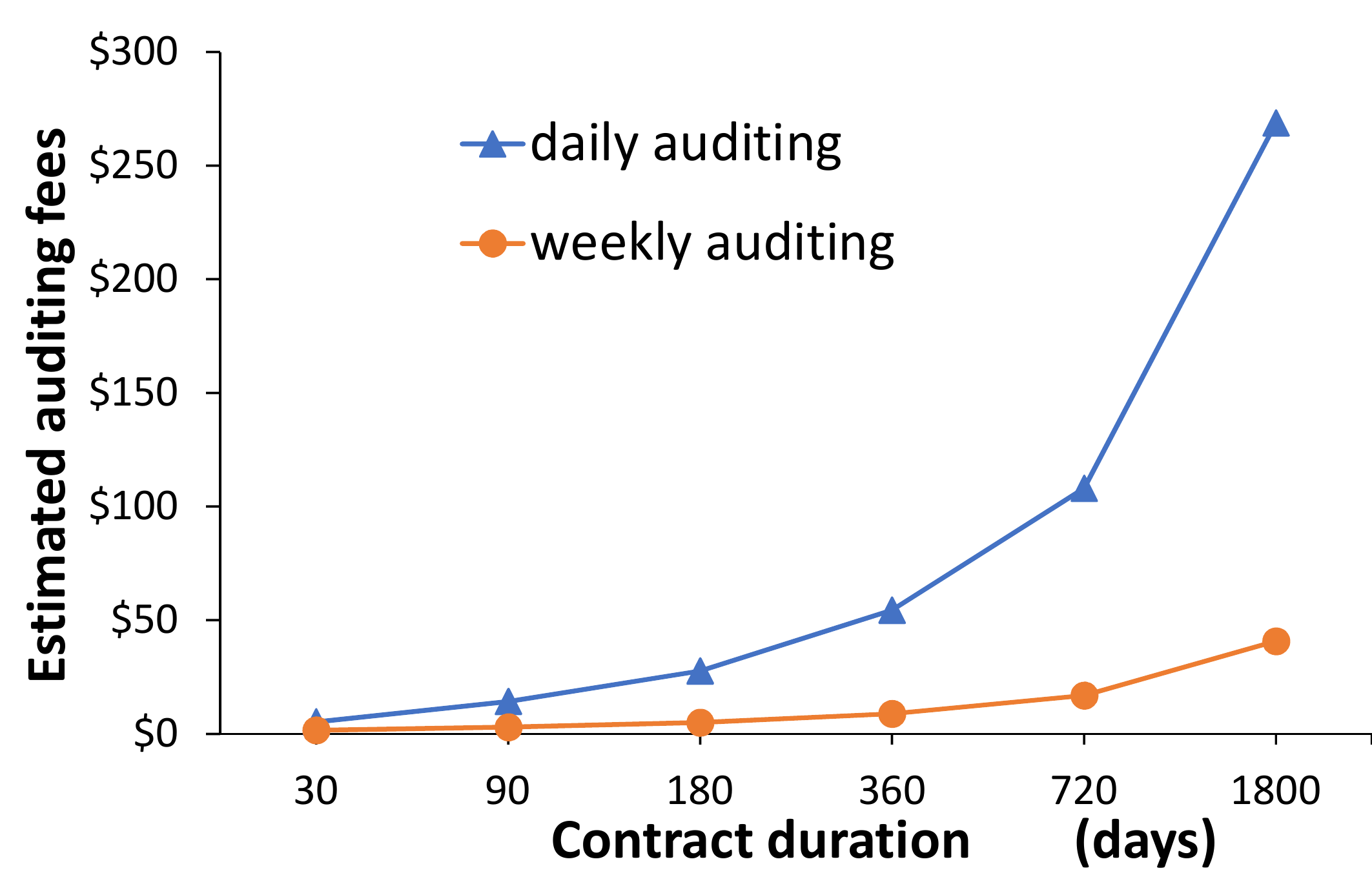}
\caption{Auditing fees from contract duration with tunable auditing frequency.}
\label{fig:auditcost}
\end{minipage}
\end{figure*}

For efficient instantiation of pairing-based cryptography, we leverage the BN256 curve ($|p| = |\mathbb{G}_{1}| = 256$ bits, $|\mathbb{G}_{2}| = 512$ bits, $|\mathbb{G}_{T}| = 1536$ bits), which is implemented in the Golang library~\cite{bn256} with optimized assembly language for elliptic-curve related operations.
%
%
Another major computational overhead, aside from elliptic curve operations, is contributed by the finite field operations over $\mathbb{Z}_{p}$.
Due to the constraints of native math/big library~\cite{golang-bignum} of Golang, we \red{strive to optimize finite field operations over $\mathbb{Z}_{p}$.}
Lastly, note that all evaluation outcomes represent the average results of 100 trials.

\subsection{On-chain efficiency and auditing cost}
\label{subsec-onchain}

The top concern for most decentralized applications attributes to the on-chain cost.
In our scenario, the on-chain cost also accounts for the majority of storage fees in our blockchain-enabled decentralized storage system.
Throughout the whole auditing procedures, the data owner needs to pay the on-chain cost for the following items.

\noindent\textbf{One-time storage cost.} 
During the pre-processing phase, our auditing protocol depends on public keys stored on the blockchain.
Figure~\ref{fig:keysize} shows the range of the one-time cost to fluctuate.
We stress, though, this cost would be no more than a few US dollars~\footnote{ETH price is 143 USD/ETH and gas cost is 5 Gwei~\cite{eth-price}, as of Apr 2020.}, irrelevant of the storage contract duration.

\noindent\textbf{Per audit cost.} 
As the ZK-SNARK framework \red{serves as our strawman solution}, we compare our \red{main solution with our extremely streamlined SNARK-based solution} at the 128 bit security level in Table~\ref{tab-exp}.
%
%
%
To achieve a relatively fair comparison, we assume the gas cost incurred by the computational overhead proportional to the computational time. 
We then adopt the ZK-SNARK verification transaction on the Ropsten Testnet~\cite{zktransaction} as the \red{baseline benchmark, which is a deployed ZK-SNARK contract on the main network}.
Our approach for extrapolation is illustrated in Fig.~\ref{fig:verify}.

We aggressively optimize the on-chain computational performance, compared to the benchmark implementation~\cite{ethersnark,googlebn256}.
With the security parameters in our experiment, it would only cost approximately $589, 000$ gases per auditing ($7.2$ ms for verification, proof size $288$ bytes, containing 3 $\mathbb{G}_{1}$ and 1 $\mathbb{G}_{T}$).
%
%
\red{To compare with standard generic SNARK-based solutions, the on-chain cost of our secure auditing protocol is $50\%$ cheaper.}
\red{To conclude, our main solution produces more succinct on-chain proofs with less time to compute.}

\noindent\textbf{Cost-effectiveness of randomness generation.} For each of the time window for challenging, the smart contract has to grab enough randomness for producing unique $C_{1}, C_{2}, r$ (48 bytes). 
The cost of such services ranges from 0.01\$~\cite{schindler2018hydrand} to 0.05\$~\cite{randao} according to our estimation, which accounts for a very small percentage of overall cost.

\noindent\textbf{Estimated annual fees.} \red{Compared to cloud storage scenarios, the most significant additional cost of our decentralized archive storage attributes to the overall on-chain cost, which is proportional to the tunable auditing frequency.}
\red{It is noteworthy that in practice, the auditing frequency is supposed to be adjusted to the order of a day (or greater), considering the blockchain latency issues and practical needs of storage guarantees.}

\red{In Fig.~\ref{fig:auditcost}, we display the relation between the auditing frequency and the overall cost when the contract duration is fixed.}
By keeping a daily auditing contract and even considering the data redundancy factors illustrated in Section~\ref{subsec-background} (e.g., 3-out-of-10 erasure coding), the annual auditing \red{on-chain cost is close to} the same level of most cloud storage providers' annual storage fees\footnote{Dropbox Business~\cite{Dropbox} offers a standard price of 150\$, as of Apr 2020.}.

\begin{figure*}[t]
	\begin{minipage}{0.33\linewidth}
		\centering
		\includegraphics[width=1.0\linewidth]{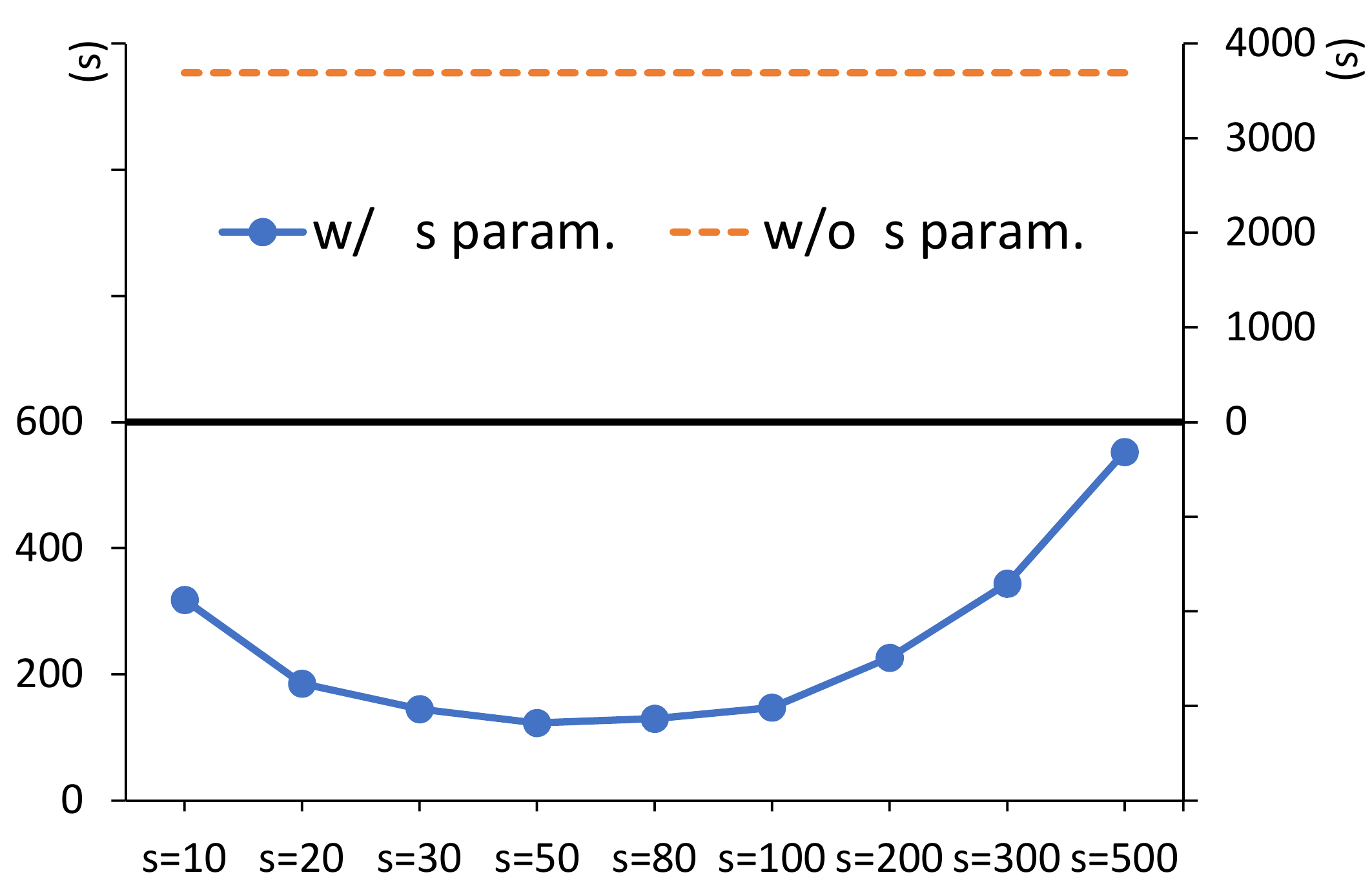}
		\captionsetup{width=0.95\textwidth}
		\caption{Time for $\mathcal{D}$ to pre-process 1 GB data with quad-core CPUs.
			When $s$ is tuned to around 50, $\mathcal{D}$ achieves the optimal time.
			Note this pre-processing time is proportional to the file size.
			\red{In the case of $s = 50$, pre-processing speed is around 35.31 MB/s}.
		}
		\label{fig:preprocess}
	\end{minipage}
	\begin{minipage}{0.33\linewidth}
		\centering
		\includegraphics[width=1.0\linewidth]{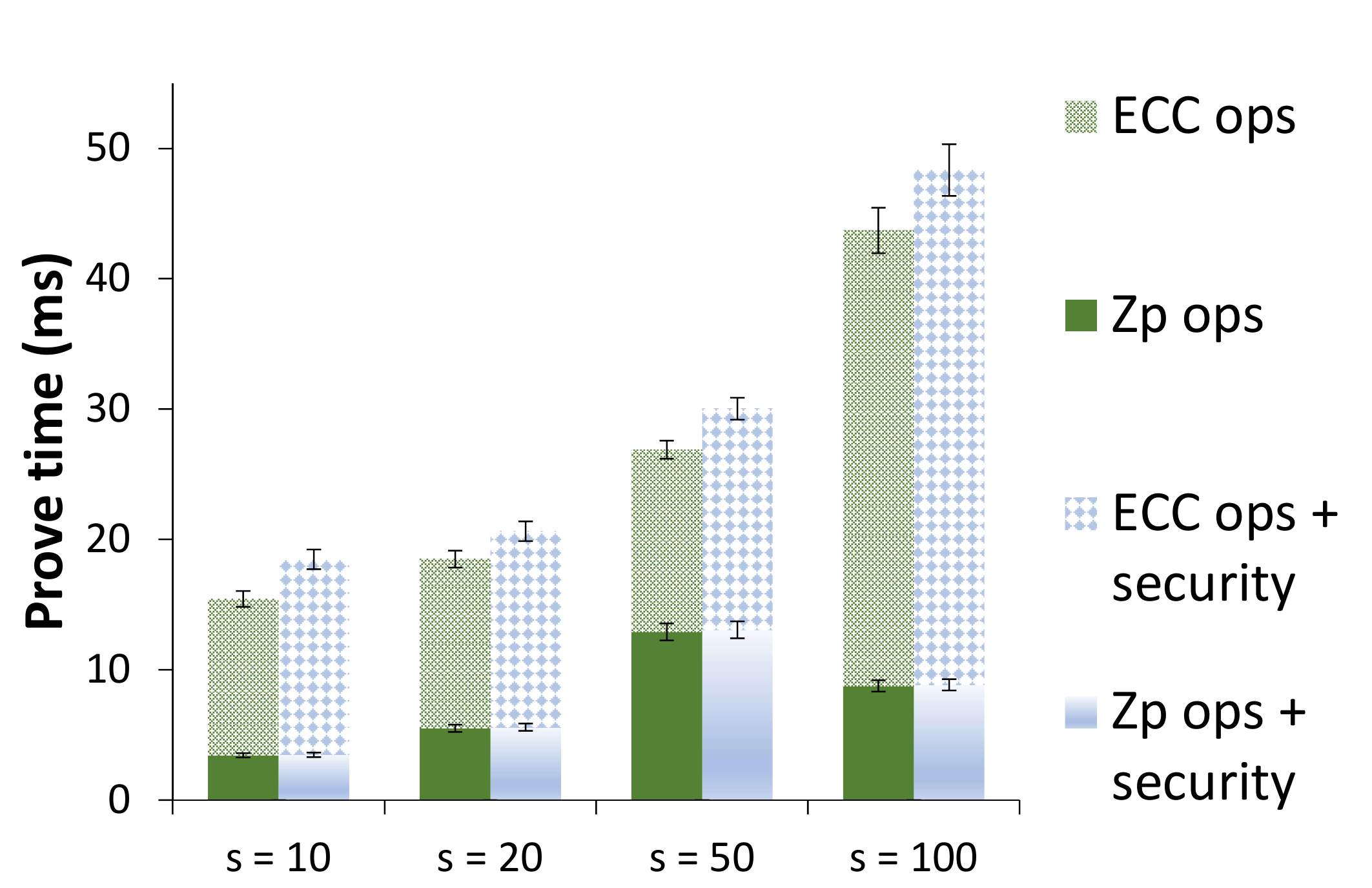}
		\captionsetup{width=0.95\textwidth}
		\caption{Time for $\mathcal{S}$ to generate a proof, $k$ = 300 (i.e., 95\% conf. when 1\% data corrupted).
			Time for $\mathbb{Z}_{p}$ operations peaks when $s$ is around 50.
			\red{Yet this amount of time for $\mathbb{Z}_{p}$ still counts as a minor role.}
			By and large, ECC operations dominate the running time.}
		\label{fig:prove}
	\end{minipage}
	\begin{minipage}{0.33\linewidth}
		\centering
		\includegraphics[width=1.0\linewidth]{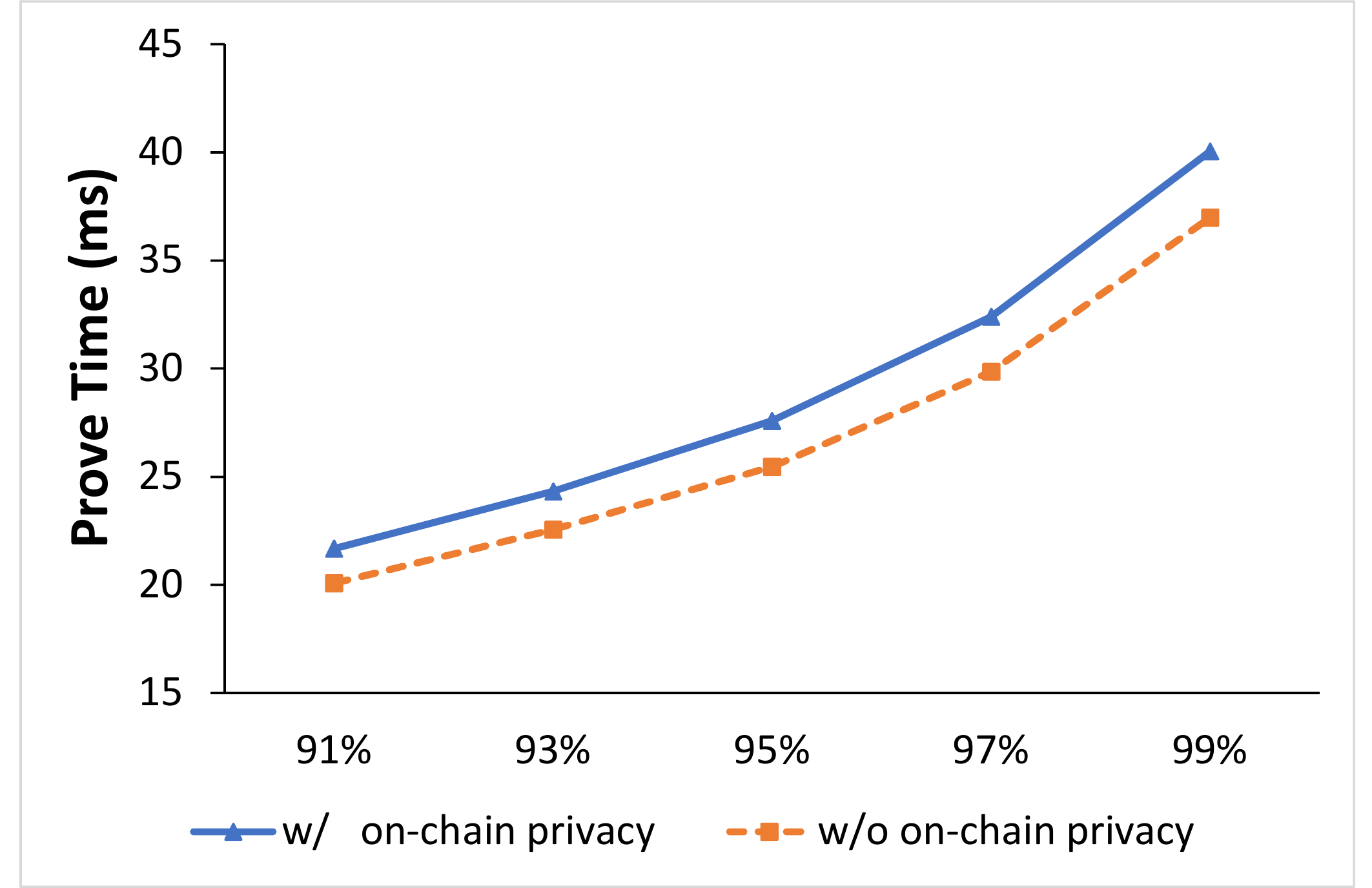}
		\captionsetup{width=0.95\textwidth}
		\caption{\red{Time for generating a proof increases, as the probability of storage guarantees goes up (when there is 1\% data corruption).
				The dotted line shows the proving process without any on-chain privacy guarantees whilest the solid line shows the opposite.}}
		\label{fig:prove-prob}
	\end{minipage}
	\vspace{-5pt}
\end{figure*}

\subsection{Off-chain overhead}
\label{subsec-convenience}
As shown in Table~\ref{tab-exp}, SNARK-based solutions incur tremendous overhead on the data owner during the trusted setup phase and on the storage provider during the proof generation phase.
While our auditing protocol would largely alleviate the overhead on both the data owner and the storage provider.

\noindent\textbf{Minimized work for data owner.} Compared to the heavy-weight SNARK-based solution, our auditing protocol only requires the data owner to invest a one-time and rather insignificant amount of time for pre-processing.
\red{In contrast, SNARK-based solutions produce public parameters of size up to hundreds of Megabytes, and require a massive amount of time even for a small piece of data (see our strawman solution evaluation of pre-processing in Table~\ref{tab-exp}).}
Compare to prior arts~\cite{shacham2008compact,wang2013privacy}, our solution largely diminish the time required, as illustrated in Fig.~\ref{fig:preprocess}.
To outsource a file of $1 GB$ size, the data owner only needs to spend $2$ minutes when setting the storage/computational parameter $s$ to $50$ on a quad-core laptop.
Notice \red{that aside from the signature generation time during the user's preprocessing period, we also count other factors such as key generation and polynomial coefficient transformation of data blocks, into preprocessing time in a broader sense.}
%

\begin{figure}
\centering
\begin{subfigure}{0.5\linewidth}
\includegraphics[width=1.2\linewidth]{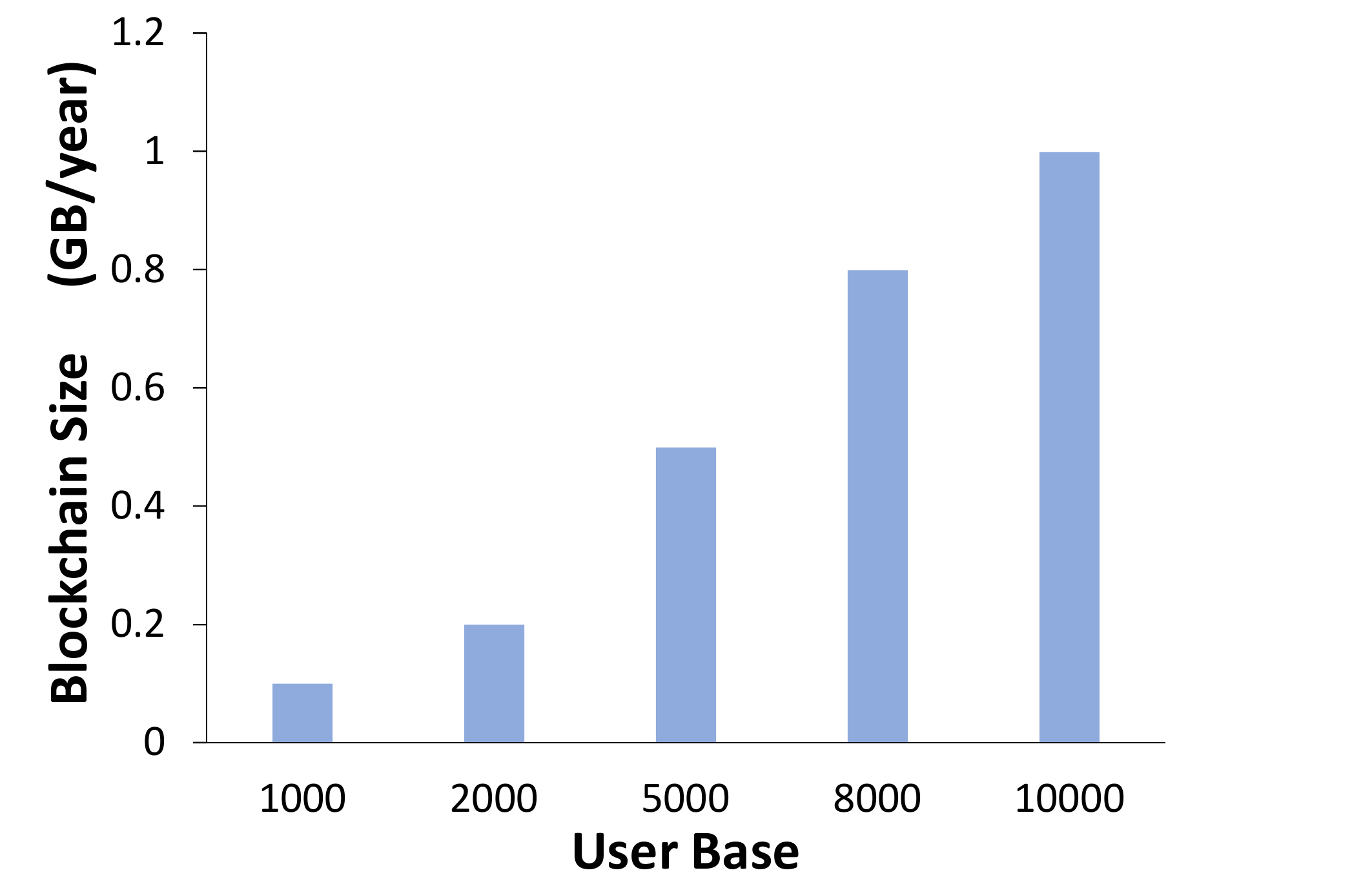}
\end{subfigure}%
\begin{subfigure}{0.5\linewidth}
\includegraphics[width=1.2\linewidth]{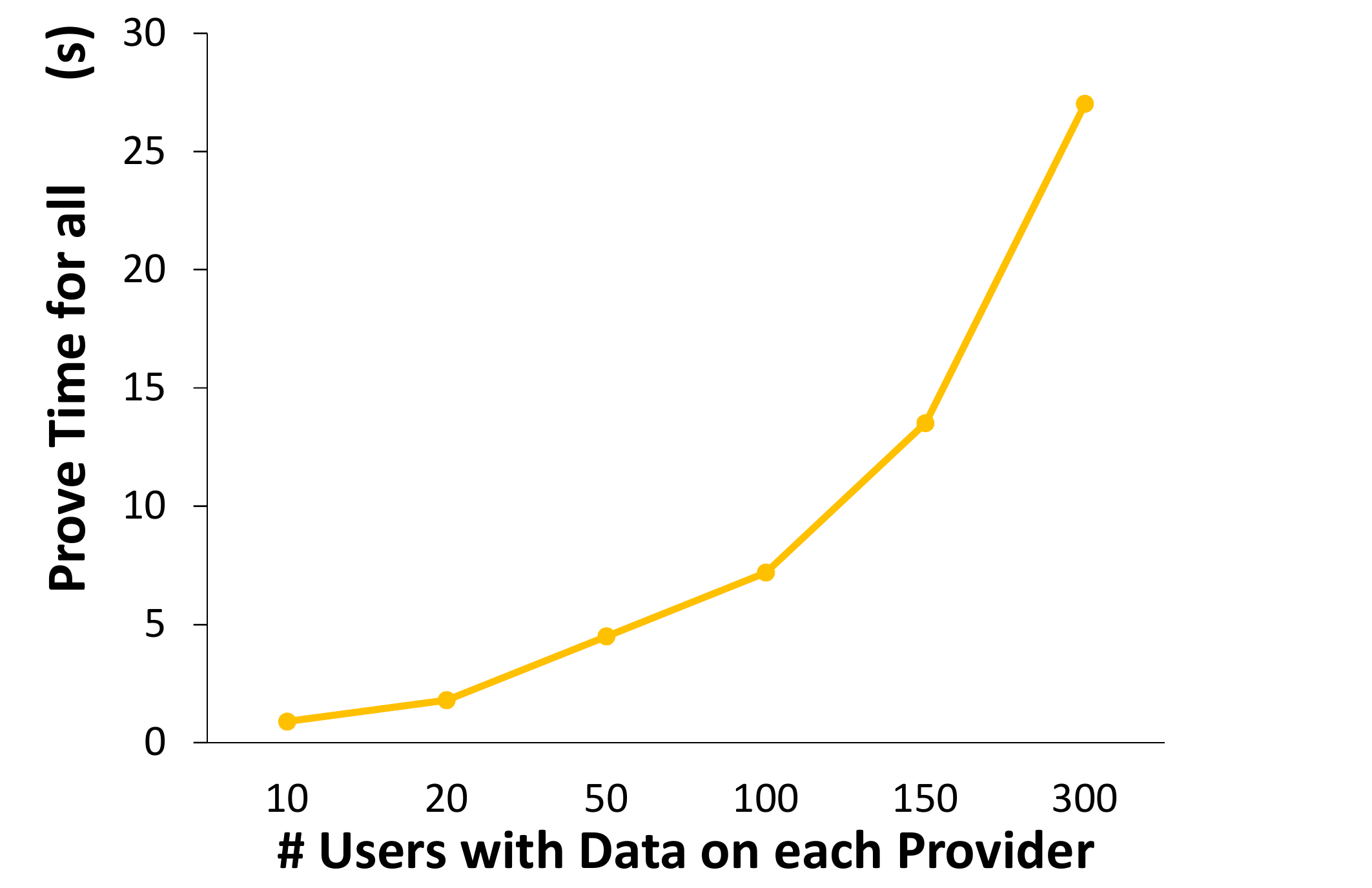}
\end{subfigure}
\caption{Annual growth of blockchain size and time for processing all contracts on each $\mathcal{D}$.
\red{The major contributing factor for both is the user base, which could be in the case of thousands}.
\red{Note that the latter is directly affected by \# of $\mathcal{D}$ (typically dozens of) with data on each $\mathcal{S}$.}}
\label{fig:scale}
\vspace{-15pt}
\end{figure}

\noindent\textbf{Efficient storage provider.} The storage providers also benefit from our auditing solutions. 
By introducing the parameter $s$, the size of extra storage taken from the storage provider is only $1/s$ of the original data size.
Although by increasing the parameter $s$, the time for generating proofs also grows, the actual results are quite gratifying, given all the cryptographic operations in $\mathbb{Z}_{p}$ and on elliptic curves.
Figure~\ref{fig:prove} displays the processing time for detailed operations, when the number of challenged blocks is set to $300$.
We stress that it is important to restrict the computational overhead on the side of storage providers.
\red{We also present a more thorough comparison, when the number of challenged blocks varies and suits for different storage guarantee levels.}
\red{In Fig.~\ref{fig:prove-prob}, we find that the time for generating a proof increases significantly, as the confidence level of storage guarantees climbs from 91\% (number of challenged blocks equal to 240) to 99\% (number of challenged blocks equal to 460), given 1\% of data tampering.}

\subsection{System-wide factors}
\label{subsec-scalability}

\noindent\textbf{Blockchain throughput.} For a single data owner to store data, the auditing cost is mostly affected by the factor of auditing frequency.
However, when the scale of user base grows, the scalability of the entire decentralized storage system is constricted by the throughput of the blockchain.
In the real-world deployment, we could use a dedicated Ethereum fork for the auditing use only.
We further assume the average block size is around 18 KB, which is closely to the average block size on Ethereum~\cite{etherscan} in the past 6 months.
Under this assumptions, the average throughput would be 2 transactions per second.
Even we take the factor of data redundancy for data spares into consideration, our system could support $5, 000$ active users at the same time with ease.
For comparison, both Storj~\cite{storjstat} and Siacoin only support a $1,000$ user storage system presently.

\noindent\textbf{Scalability.} As indicated in the left sub-figure of Fig.~\ref{fig:scale}, the estimated annual growth of the blockchain size would be at a slower place, compared to \red{a daily growth of around $128$ MB~\cite{eth-tx} for} the present Ethereum platform.
From our collected data on the Siacoin and Storj network, we discover that in a system with $1,000$ users, on average a storage provider may have to store data from approximately $30$ users.
Though our auditing protocol natively supports the batch auditing~\cite{wang2013privacy}, the proof generation time still grows proportionally with the number of public keys.
The sub-figure on the right side of Figure~\ref{fig:scale} displays the overall proving time required for each storage provider when the number of users who store data on them increases progressively, assuming a linear regression model. 
In particular, it may cost the storage provider approximately $20$ seconds to complete all proof generation procedures when the number of users is $5,000$.
Yet we argue this amount of time is tolerable, as the latency (avg. confirmation time) on the asynchronized blockchain costs a similar amount of time.

\section{Related work}
\label{relatedwork}
Numerous research has been conducted upon proof of storage in the past decade.
The most straightforward auditing scheme is applying the standard hash function or message authentication codes (MAC) to check the data integrity at an untrusted location.
Despite the computational efficiency, this scheme does not scale due to the inconvenience that the verifier has to re-compute the result with the same data input.
Also, it cannot support unlimited times of challenges.
Consequently, most deployed solutions construct a Merkle tree over the data to reduce the communication and verification overhead.

In the settings of cloud storage auditing, most designs leverage the public key cryptography~\cite{shacham2008compact,ateniese2007provable}, which aggregates the authenticators of blocks and thus produces short proofs.
%
%
%
Later on, the developments in the field of public verification~\cite{shacham2008compact,wang2009enabling}, efficiency improvements~\cite{xu2012towardsefficientpor,yuan2013proofs} and dynamism~\cite{Erway2009DPD,wang2009enabling,shi2013practical} have led to a renewed interest in the literature of storage auditing and its further application in the cloud storage settings.
However, these prior auditing designs cannot satisfactorily address the unique challenges in the paradigm of decentralized archive storage.

\section{Conclusion}
\label{sec-conclusion}
Envisioning a future of alternative decentralized storage, we propose a pragmatic and secure auditing protocol with strong on-chain privacy guarantees and least amount of on-chain overhead. 
%
%

\bibliographystyle{ieeetr}
\bibliography{paper}

\begin{thebibliography}{10}

\bibitem{cohen2003incentives}
B.~Cohen, ``Incentives build robustness in bittorrent.''
  \url{http://bittorrent.org/bittorrentecon.pdf}, 2003.

\bibitem{benet2014ipfs}
J.~Benet, ``{IPFS} --- content addressed, versioned, {P2P} file system,'' {\em
  CoRR}, vol.~abs/1407.3561, 2014.

\bibitem{nakamoto2008bitcoin}
S.~Nakamoto, ``Bitcoin: A peer-to-peer electronic cash system,'' 2008.

\bibitem{wood2014ethereum}
G.~Wood, ``Ethereum: A secure decentralised generalised transaction ledger,''
  {\em Ethereum project yellow paper}, 2014.

\bibitem{finck2018blockchains}
M.~Finck, ``Blockchains and data protection in the european union,'' {\em Eur.
  Data Prot. L. Rev}, vol.~4, p.~17, 2018.

\bibitem{GDPRpositionpaper}
N.~Eichler, S.~Jongerius, G.~McMullen, O.~Naegele, L.~Steininger, and
  K.~Wagner, ``Blockchain, data protection, and the gdpr,'' 2018.

\bibitem{storjstat}
``Storjstat.'' \url{https://storjstat.com/}, 2019.

\bibitem{siacoin2014}
L.~C. David~Vorick, ``Sia: Simple decentralized storage,'' 2014.

\bibitem{filecoin2017}
P.~Labs, ``Filecoin: A decentralized storage network,'' 2017.

\bibitem{us-smartcontract-court}
{Scott Kimpel and Christopher Adcock}, ``The state of smart contract
  legislation.''
  \url{https://www.blockchainlegalresource.com/2018/09/state-smart-contract-legislation},
  2018.

\bibitem{yahoo-chinasupremecourt}
{Mark Emem}, ``Blockchain records will now be accepted as legal evidence,
  china’s supreme court rules.''
  \url{https://finance.yahoo.com/news/blockchain-records-now-accepted-legal-121050674.html},
  2018.

\bibitem{storj2018}
S.~Labs, ``Storj: A decentralized cloud storage network framework,'' 2018.

\bibitem{zkcsp}
M.~Campanelli, R.~Gennaro, S.~Goldfeder, and L.~Nizzardo, ``Zero-knowledge
  contingent payments revisited: Attacks and payments for services,'' in {\em
  Proc. of ACM CCS}, 2017.

\bibitem{hawk}
A.~E. Kosba, A.~Miller, E.~Shi, Z.~Wen, and C.~Papamanthou, ``Hawk: The
  blockchain model of cryptography and privacy-preserving smart contracts,'' in
  {\em Proc. of IEEE S{\&}P}, 2016.

\bibitem{bowers2009proofs}
K.~D. Bowers, A.~Juels, and A.~Oprea, ``Proofs of retrievability: theory and
  implementation,'' in {\em Proc. of ACM CCS Workshop}, 2009.

\bibitem{stoica2001chord}
I.~Stoica, R.~T. Morris, D.~Liben{-}Nowell, D.~R. Karger, M.~F. Kaashoek,
  F.~Dabek, and H.~Balakrishnan, ``Chord: a scalable peer-to-peer lookup
  protocol for internet applications,'' {\em {IEEE/ACM} Trans. Netw}, vol.~11,
  no.~1, pp.~17--32, 2003.

\bibitem{PoRep2018}
B.~Fisch, ``Tight proofs of space and replication,'' in {\em Proc. of
  EUROCRYPT}, 2019.

\bibitem{rabin1983beacon}
M.~O. Rabin, ``Transaction protection by beacons,'' {\em J. Comput. Syst. Sci},
  vol.~27, no.~2, pp.~256--267, 1983.

\bibitem{parno2013pinocchio}
B.~Parno, J.~Howell, C.~Gentry, and M.~Raykova, ``Pinocchio: Nearly practical
  verifiable computation,'' in {\em Proc. of IEEE S{\&}P}, 2013.

\bibitem{ben2013snarks4C}
E.~Ben{-}Sasson, A.~Chiesa, D.~Genkin, E.~Tromer, and M.~Virza, ``Snarks for
  {C:} verifying program executions succinctly and in zero knowledge,'' in {\em
  Proc. of CRYPTO}, 2013.

\bibitem{ben2014succinct}
E.~Ben{-}Sasson, A.~Chiesa, E.~Tromer, and M.~Virza, ``Succinct non-interactive
  zero knowledge for a von neumann architecture,'' in {\em Proc. of USENIX
  Security}, 2014.

\bibitem{groth2016size}
J.~Groth, ``On the size of pairing-based non-interactive arguments,'' in {\em
  Proc. of EUROCRYPT}, 2016.

\bibitem{wang2009ensuring}
C.~Wang, Q.~Wang, K.~Ren, and W.~Lou, ``Ensuring data storage security in cloud
  computing,'' in {\em Proc. of IEEE/ACM IWQoS}, 2009.

\bibitem{wang2013privacy}
C.~Wang, Q.~Wang, K.~Ren, and W.~Lou, ``Privacy-preserving public auditing for
  data storage security in cloud computing,'' in {\em Proc. of IEEE INFOCOM},
  2010.

\bibitem{shacham2008compact}
H.~Shacham and B.~Waters, ``Compact proofs of retrievability,'' {\em J.
  Cryptology}, vol.~26, no.~3, pp.~442--483, 2013.

\bibitem{xu2012towardsefficientpor}
J.~Xu and E.~Chang, ``Towards efficient proofs of retrievability,'' in {\em
  Proc. of ACM ASIACCS}, 2012.

\bibitem{yuan2013proofs}
J.~Yuan and S.~Yu, ``Proofs of retrievability with public verifiability and
  constant communication cost in cloud,'' in {\em Proc. of ACM ASIACCS-SCC},
  2013.

\bibitem{schnorr91}
C.~Schnorr, ``Efficient signature generation by smart cards,'' {\em J.
  Cryptology}, vol.~4, no.~3, pp.~161--174, 1991.

\bibitem{KateZG10}
A.~Kate, G.~M. Zaverucha, and I.~Goldberg, ``Constant-size commitments to
  polynomials and their applications,'' in {\em Proc. of ASIACRYPT}, 2010.

\bibitem{libert2016functional}
B.~Libert, S.~C. Ramanna, and M.~Yung, ``Functional commitment schemes: From
  polynomial commitments to pairing-based accumulators from simple
  assumptions,'' in {\em Proc. of ICALP}, 2016.

\bibitem{eclipse-bitcoin}
E.~Heilman, A.~Kendler, A.~Zohar, and S.~Goldberg, ``Eclipse attacks on
  bitcoin's peer-to-peer network,'' in {\em Proc. of USENIX Security}, 2015.

\bibitem{eclipse-ethereum}
Y.~Marcus, E.~Heilman, and S.~Goldberg, ``Low-resource eclipse attacks on
  ethereum's peer-to-peer network.'' Cryptology ePrint Archive, Report
  2018/236, 2018.
\newblock \url{https://eprint.iacr.org/2018/236}.

\bibitem{syta2017RandHerd}
E.~Syta, P.~Jovanovic, E.~Kokoris{-}Kogias, N.~Gailly, L.~Gasser, I.~Khoffi,
  M.~J. Fischer, and B.~Ford, ``Scalable bias-resistant distributed
  randomness,'' in {\em Proc. of IEEE S{\&}P}, 2017.

\bibitem{schindler2018hydrand}
P.~Schindler, A.~Judmayer, N.~Stifter, and E.~Weippl, ``Hydrand: Practical
  continuous distributed randomness.'' Cryptology ePrint Archive, Report
  2018/319, 2018.
\newblock \url{https://eprint.iacr.org/2018/319}.

\bibitem{randao}
``Randao: Verifiable random number generation.'' \url{https://www.randao.org/},
  2019.

\bibitem{chatterjee2019probabilistic}
K.~Chatterjee, A.~K. Goharshady, and A.~Pourdamghani, ``Probabilistic smart
  contracts: Secure randomness on the blockchain,'' in {\em Proc. of IEEE
  ICBC}, 2019.

\bibitem{bunz2017proofs}
D.~Boneh, J.~Bonneau, B.~B{\"{u}}nz, and B.~Fisch, ``Verifiable delay
  functions,'' in {\em Proc. of CRYPTO}, 2018.

\bibitem{temtum}
``Temtum: White paper.'' \url{https://temtum.com/whitepaper/}, 2019.

\bibitem{boneh2004short}
D.~Boneh and X.~Boyen, ``Short signatures without random oracles,'' in {\em
  Proc. of EUROCRYPT}, 2004.

\bibitem{ateniese2007provable}
G.~Ateniese, R.~C. Burns, R.~Curtmola, J.~Herring, L.~Kissner, Z.~N.~J.
  Peterson, and D.~X. Song, ``Provable data possession at untrusted stores,''
  in {\em Proc. of ACM CCS}, 2007.

\bibitem{douceur2002sybil}
J.~R. Douceur, ``The sybil attack,'' in {\em Proc. of IPTPS}, 2002.

\bibitem{bellman}
Zcash, ``Github zk-snark library.''
  \url{https://github.com/zcash/librustzcash/tree/master/bellman}, 2019.

\bibitem{xie2019libra}
T.~Xie, J.~Zhang, Y.~Zhang, C.~Papamanthou, and D.~Song, ``Libra: Succinct
  zero-knowledge proofs with optimal prover computation,'' in {\em Proc. of
  CRYPTO}, 2019.

\bibitem{rsa-pos}
G.~Ateniese, A.~Faonio, and S.~Kamara, ``Leakage-resilient identification
  schemes from zero-knowledge proofs of storage,'' in {\em Proc. of IMACC},
  2015.

\bibitem{cramer1996modular}
C.~Ronald, ``Modular design of secure yet practical cryptographic protocols,''
  {\em Ph. D. Thesis, CWI and University of Amsterdam}, 1996.

\bibitem{tahoe}
``Tahoe-lafs.'' \url{https://tahoe-lafs.org/trac/tahoe-lafs}, 2019.

\bibitem{ewasm}
``Ethereum flavored webassembly (ewasm).''
  \url{https://github.com/ewasm/design}, 2019.

\bibitem{bn256}
Cloudflare, ``Package bn256 implements a particular bilinear group at the
  128-bit security level.'' \url{https://github.com/cloudflare/bn256}, 2019.

\bibitem{golang-bignum}
``math/big: implement recursive division algorithm.''
  \url{https://github.com/golang/go/issues/21960}, 2017.

\bibitem{eth-price}
``Recommended gas prices in gwei.'' \url{https://www.ethgasstation.info/},
  2019.

\bibitem{zktransaction}
``Etherscan transaction details.''
  \url{https://ropsten.etherscan.io/tx/0x15e7f5ad316807ba16fe669a07137a5148973235738ac424d5b70f89ae7625e3},
  2017.

\bibitem{ethersnark}
``Snark test in solidity.''
  \url{https://gist.github.com/chriseth/f9be9d9391efc5beb9704255a8e2989d},
  2017.

\bibitem{googlebn256}
``Bn256 pre-compiled contract in go-ethereum.''
  \url{https://github.com/ethereum/go-ethereum/blob/master/crypto/bn256/google/bn256.go},
  2017.

\bibitem{Dropbox}
``Dropbox pricing.'' \url{https://www.dropbox.com/business/pricing}, 2019.

\bibitem{etherscan}
Ethereum, ``Etherscan.'' \url{https://etherscan.io/}, 2019.

\bibitem{eth-tx}
``Ethereum transaction history.'' \url{https://etherscan.io/chart/tx}, 2019.

\bibitem{wang2009enabling}
Q.~Wang, C.~Wang, J.~Li, K.~Ren, and W.~Lou, ``Enabling public verifiability
  and data dynamics for storage security in cloud computing,'' in {\em Proc. of
  ESORICS}, 2009.

\bibitem{Erway2009DPD}
C.~C. Erway, A.~K{\"{u}}p{\c{c}}{\"{u}}, C.~Papamanthou, and R.~Tamassia,
  ``Dynamic provable data possession,'' in {\em Proc. of ACM CCS}, 2009.

\bibitem{shi2013practical}
E.~Shi, E.~Stefanov, and C.~Papamanthou, ``Practical dynamic proofs of
  retrievability,'' in {\em Proc. of ACM CCS}, 2013.

\end{thebibliography}

\end{document}